\documentclass[smallextended]{svjour3}
\usepackage{appendix}
\usepackage{amssymb}
\usepackage{amsmath}
\usepackage{graphicx}
\usepackage{bm}
\def\Xint#1{\mathchoice
   {\XXint\displaystyle\textstyle{#1}}%
   {\XXint\textstyle\scriptstyle{#1}}%
   {\XXint\scriptstyle\scriptscriptstyle{#1}}%
   {\XXint\scriptscriptstyle\scriptscriptstyle{#1}}%
   \!\int}
\def\XXint#1#2#3{{\setbox0=\hbox{$#1{#2#3}{\int}$}
     \vcenter{\hbox{$#2#3$}}\kern-.5\wd0}}

\def\dashint{\Xint-}

\begin{document}

\pagestyle{plain}

\title{
\begin{minipage}[t]{7.0in}
\scriptsize
\begin{quote}
\leftline{{\it J. Low Temp. Phys.}, in press.}
\leftline {\rm arXiv:1609.03538}
\end{quote}
\end{minipage}
\medskip
Stoner-type theory of Magnetism in Silicon MOSFETs
}
\author{D. I. Golosov}
%\email{Denis.Golosov@biu.ac.il}
%\affiliation{Department of Physics and the Resnick Institute, Bar-Ilan 
%University, Ramat-Gan 52900, Israel.}

\institute{D. I. Golosov \at 
Department of Physics and the Resnick Institute, Bar-Ilan 
University, Ramat-Gan 52900, Israel\\
\email{Denis.Golosov@biu.ac.il}}

\date{\today}

\maketitle

\begin{abstract}
We consider quasi-two-dimensional gas of electrons in a typical Si-MOSFET, 
assuming repulsive contact interaction between electrons. Magnetisation and 
susceptibility are evaluated within the mean-field approach. 
Finite thickness of the inversion layer results in an interaction-induced 
electron
wave function change, not found in both purely two-dimensional and 
three-dimensional (bulk) cases.  Taking this self-consistent change
into account leads to an increased susceptibility and ultimately to a 
ferromagnetic transition deep in the high-density metallic regime.
We further find that in the paramagnetic state, magnetisation increases 
sublinearly with increasing  in-plane magnetic field.
In the opposite limit of low carrier densities, the effects of long-range 
interaction become important and can be included phenomenologically via
bandwidth renormalisation. Our treatment then suggests that with decreasing 
density, the metal-insulator transition is preceded by a ferromagnetic 
instability. Results are discussed in the context of
the available experimental data, and arguments for the validity of our 
mean-field scheme are presented.  
%\end{abstract}

\end{abstract}
%\maketitle

\section{Introduction}
\label{sec:intro}

 Silicon 
metal-oxide-semiconductor field-effect transistors (Si-MOSFETs) have been 
in the focus of an extensive research effort throughout  the ongoing studies 
of the properties of low-dimensional electron systems.
Fifty years ago, electrons in the Si-MOSFET inversion layers\cite{Fowler66} 
were among the
first experimental realisations of 2-dimensional (2D) electron gas (2DEG) 
\cite{Ando82}. Some thirty years later, they yielded the first example of a 
metal-insulator transition (MIT) in a 2D 
system\cite{Kravchenko94,Kravchenko04}. While the full 
understanding of this phenomenon is yet to be achieved, a remarkable progress
toward this goal has been made both experimentally and 
theoretically\cite{Spivak10}. In particular, attention was paid to the close 
interplay between spin and charge degrees
of freedom, as exemplified by strong positive magnetoresistance in a parallel
magnetic field (when orbital effects are negligible)\cite{Kravchenko04,epl12}. 
This allows, for example,
to use an electric current for manipulating the spin density in restricted 
geometries\cite{prb13}, which appears relevant in the general context 
of spintronics. 

As opposed to the magnetotransport, measuring the magnetic properties of the 
2DEG presents formidable experimental 
difficulties\cite{Kravchenko06,Kravchenko06-2,Pudalov15}. In the case of 
Si-MOSFETs, 
such measurements are necessarily indirect, and both accuracy and 
interpretation of the results can and should be questioned. Nevertheless,
it was established that the low-field magnetic susceptibility in the metallic 
state increases when the carrier density (controlled by a voltage applied to 
the metallic gate) is decreased toward the MIT. It is not yet reliably 
verified whether this increase is finite\cite{Reznikov03}, or a 
ferromagnetic transition 
takes place in the vicinity of the MIT\cite{Kravchenko06,Kravchenko06-2}; 
in addition, 
evidence of magnetic inhomogeneities has been reported recently for low 
densities\cite{Teneh12}. 
An important 
theoretical study suggests
a divergence of the electron effective mass and hence of the susceptibility at 
the MIT without an associated magnetic transition\cite{Finkelstein05}.

It should be noted that the possibility of ferromagnetism in a 2DEG is a 
fascinating subject in itself, originally suggested on the basis of numerical
investigations\cite{Tanatar89}. While this suggestion finds further support
in some subsequent numerical work \cite{Attaccalite02}, others\cite{Senatore09}
do not find any critical behaviour of susceptibility in a low density 
two-valley 2DEG (the latter, as appropriate for a Si-\{100\} MOSFET).  
It was also noted that in reality, the 
inversion layer has a finite thickness (which increases for smaller carrier 
densities), resulting in a quasi-2DEG (as opposed to a strictly 2D case).
This was taken into account by including the appropriate 
formfactors\cite{Ando82} into diagrammatic 
summations\cite{DasSarma05,ZhangDasSarma2005} 
and Monte-Carlo numerical calculations \cite{Senatore05,Asgari2006}. 
Orbital effects of 
the in-plane magnetic field were invoked as well \cite{Tutuc03}.

Thus, the magnetic properties of 2DEG in Si-MOSFETs attract considerable 
attention from both theorists and experimentalists. It is therefore somewhat
surprising that a systematic Stoner-type mean field treatment has not been
carried out for this case. This is probably due to the fact that in a 2DEG at
low densities (in the vicinity of the MIT), the dominant role is played by 
the long-range Coulomb correlations, whereas the Stoner approach emphasises 
the {\it local} mean field, arising from the short-range (or on-site) 
repulsion.  

In agreement with Ref. \cite{Finkelstein05}, a comprehensive recent 
review\cite{Dolgopolov15} of
experimental data finds a pronounced renormalisation of quasiparticle
band on approaching the MIT from the high-density metallic phase. 
Phenomenologically, the data correspond to a non-interacting 2DEG with a 
bandwidth vanishing at the MIT. We argue that the effects of the short-range
interaction  likely become important in this case, drastically modifying the
magnetic properties of the system and leading to a ferromagnetic transition.  
This surely holds also in the opposite case of high carrier density (deep inside
the metallic phase): there, the on-site repulsion provides the dominant 
contribution to magnetic susceptibility, which increases with density.
Due to the restricted geometry of electron motion in  MOSFETs (finite layer
thickness), the mean field theory takes on a somewhat unusual form as opposed
to purely 2D or 3D cases. In addition to Zeeman-like energy 
shifts under a combined effect of interaction and external field, one must take
into account changes in the carrier wave functions. This effect, which
was not included in previous treatments, leads to a further increase in 
susceptibility. This opens an intriguing possibility of a ferromagnetic 
transition
in the region where the interaction is still not too strong, and hence the mean 
field approach is qualitatively valid. The latter should be contrasted with the
well-known failure of mean field theory for the two-dimensional Hubbard model,
where even in the case of infinite on-site repulsion ferromagnetism may arise
only in a restricted range of values of the carrier 
density\cite{Dombre,Edwards,Wurth} 
(although the ferromagnetic region of the phase diagram is broadened once the
allowance is made for further-neighbour hopping\cite{Hlubina99} and for partial 
spin-polarisation in the ferromagnetic state\cite{Becca01}). On the
other hand, we note recent results\cite{Conduit2013} on 2D atomic gases with 
short-range repulsion,
suggesting that mean field theory may be {\it overestimating} the interaction
strength required for ferromagnetism.  

For the purposes of the present study, it is obviously important to 
adequately estimate the 
strength of
short-range interaction. While a recent article\cite{Cococcioni2010} suggests 
that the on-site repulsion is 
of order $U_{on-site}\approx 3$ eV, this is likely to be an over-estimate, 
especially since the Wannier function in silicon can be expected to spread 
over several lattice sites. On the other hand, we wish to write the 
short-range interaction  ${\cal U}_{3D}$ for our continuum description
in the form of a contact repulsion (or equivalently an $s$-wave scattering),
\begin{equation}
{\cal U}_{3D}=U \delta (x-x')\delta (y-y')\delta (z-z')\,.
\label{eq:contact3D}
\end{equation}
Interaction constant $U$ includes contributions from those neighbouring 
sites $j$ on the underlying  
discrete lattice where the wavefunction overlap with a given site $i$  
(or equivalently, the
off-site repulsion $U_{ij}$) is non-negligible\footnote{
While any $U_{ij}\neq 0$ with $i\neq j$ would also lead to
an interaction between same-spin carriers, it obviously cannot give rise to
an $s$-wave repulsion between these. The effects of $p$-wave and higher 
harmonics can be expected to be weak and will be neglected.}:
%\cite{samespin}:
\begin{equation}
U \sim a^3\sum_j U_{ij}\,,\,\,\,\,U_{ii}\equiv U_{on-site}\,,
\label{eq:Uest}
\end{equation} 
with 
$a \approx 5.43$ \AA ~the lattice
period.   Taking the simple cubic lattice as an example, we see that 
the combined effect of a rather more realistic $U_{on-site} \sim 0.75$ eV, 
the nearest neighbour $U_{ij} \sim 0.25$ eV and next-nearest neighbour 
repulsion of $U_{ij} \sim 0.1$ eV is the same as that of a $U_{on-site}= 3$ eV 
acting alone. 
Given the apparent absence of reliable {\it ab initio} data for $U_{ij}$, 
we will be using the latter estimate henceforth.

The outline of the present paper is as follows. The model and the mean field
scheme are introduced in Sect. \ref{sec:mf}. In the following section, we
analyse the mean field solution in the low- to moderate density range, 
where only one transverse level is occupied, discussing the emergent behaviour 
and also comparing it to a simple variational result. As explained above, 
when approaching the MIT one has to take into account the bandwidth 
renormalisation due to the long-range interactions (Sect. \ref{sec:dolgo}).
In the opposite regime of large densities, a proper description implies
filling multiple transverse levels, as described in Sect. \ref{sec:metal}. 
We note that our results suggest a possibility of ferromagnetism in both 
cases. The field dependence of magnetisation in the paramagnetic phase is 
discussed in Sect. \ref{sec:field}, and the concluding discussion is relegated
to Sect. \ref{sec:conclu}. 
Our analysis relies on a conjecture that the Stoner approach remains 
relevant in a 2D system down to sufficiently low densities. Arguments 
to this effect 
are given in the Appendix. 

Early preliminary results were reported in Ref. \cite{jmmm}.

\section{Si-MOSFET inversion layer, and the mean field scheme}
\label{sec:mf}

Here, we generalise the familiar mean-field description\cite{Stern72} of 
an $n$-doped 
Si inversion layer, taking into account the short-range electron-electron 
repulsion and allowing for the presence of an applied magnetic field.

\begin{figure} \sidecaption
\includegraphics[width=.49\textwidth]{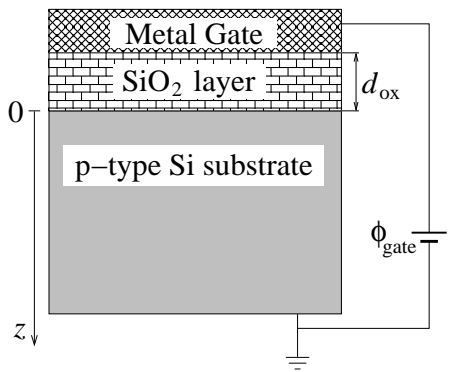}
\caption{\label{fig:MOSFET} Schematic view of a Silicon MOSFET.
}
\end{figure}

In a Si-MOSFET, a quasi-2D conducting layer is formed on the surface of bulk
silicon, and the spectral properties of the carriers depend on the 
crystallographic orientation of this surface. While this is not expected to  
affect our results at the qualitative level, we consider the case of a 
$\{1 0 0 \}$ surface.
%, where two degenerate valleys are formed in the 
%conduction band. 
When a 
sufficiently large positive voltage  $\phi_{gate}$ is applied to the metallic 
gate
(which is separated from Si by an oxide layer, see schematics in Fig. 
\ref{fig:MOSFET}), 
%these 
conduction band valleys
dip below the Fermi level. The latter is fixed at the
top of the valence band of the bulk Si, which we will use as a zero of energy,
$E_v=0$.
An adequate description of electrostatics cannot be achieved without taking 
into account the impurities present in the bulk. Again, it is expected that
the details are unimportant and we assume the presence of a single acceptor
level at $E=0$ (more precisely, at a negligible positive $E$), with the 
volume density of acceptors $N_A$. When a small positive voltage $\phi_{gate}$ 
is applied
to the gate, a {\it depletion layer} of variable width $z_d$ is formed near 
the surface. Within this layer (at $z<z_d$, assuming $z=0$ at the surface), the
acceptor states are occupied by electrons, whereas the bottom of the 
conduction band decreases linearly from its bulk value $E_c$ to a 
(variable) value $E_{cs}$ achieved at $z=0$ (triangular potential; note that 
we consider the case of zero temperature). Equivalently, the electrostatic
potential $\phi(z)$ increases linearly from its constant value 
(which we choose as $\phi=0$) within the bulk ({\it i.e.,} 
everywhere at $z>z_d$). 
With increasing 
$\phi_{gate}$, the value of $E_{cs}$ becomes negative; at a certain point 
thereafter, the lowest electronic states in the quantum-mechanical potential
well formed by the bent conduction-band bottom (see below) drop below $E=0$ 
and the quasi-2D electron gas (Q2DEG) with two-dimensional carrier 
density $n$ is formed. Throughout, the value of $E_{cs}$ is self-consistently
determined by a condition
\begin{equation}
\phi(z=0) \equiv -\frac{1}{e}(E_{cs}-E_c) =-\frac{1}{C}Q_{gate}+\phi_{gate}\,,
\label{eq:phigate}
\end{equation}
where $-e$ is the electron charge, and  $Q_{gate}$ is the (positive) surface 
charge density
at the gate, which is exactly compensated by the induced charges in the
semiconductor: $Q_{gate}=e(n+N_A z_d)$. $C$ is the capacitance per unit area of
the oxide layer,
\[
C=4\pi \epsilon_{ox}/d_{ox}
\]
where $\epsilon_{ox}$ and $d_{ox}$ are the dielectric constant of SiO$_2$ and the
layer thickness. We will be interested in the case where
$n$ exceeds the critical value $n_0$ corresponding to the MIT. In this
regime, $n$ is of the same order or larger than the two-dimensional density 
of depletion layer
charge $N_A z_d$, and the potential felt by the mobile carriers
can no longer be approximated by a triangular one; instead, we must solve a 
self-consistent
Poisson equation, which for $0<z<z_d$ takes form:
\begin{equation}
\epsilon \frac{d^2 \phi(z)}{dz^2}=
4\pi e \left[ N_A+ \sum_{l,a,\alpha}n_{l,a,\alpha} \psi^2_{l,a,\alpha}(z)\right]
, 
\label{eq:Poi}
\end{equation}
where $\epsilon$ is the static dielectric constant of Si.
The charge density on the r.h.s. includes  contributions from acceptors and 
from the Q2DEG carriers; the latter are subdivided according to the number 
of the 
corresponding transverse-motion level $a=0,1,2,...$ within the ``ladder'' $l$ 
(with $l=0,1$, see below), and the spin index 
$\alpha=\uparrow,\downarrow$.  
The corresponding 2D carrier densities and  wave functions of 
transverse motion are denoted  $n_{l,a,\alpha}$ and $\psi_{l,a,\alpha}$ 
(with $\int_0^\infty \psi^2_{l,a,\alpha}(z) dz=1$). The net densities of 
spin-up and -down electrons will be denoted by $n_\alpha$, so that
\begin{equation}
n_\alpha=\sum_{l,a} n_{l,a,\alpha}\,,\,\, 
n=n_\uparrow+n_\downarrow\,,\,\, 
M=\frac{n_\uparrow- n_\downarrow}{2}\,,
\label{eq:nm}
\end{equation}
where $M$ is magnetisation density in units of the Bohr magnetone. We will be
interested in the effects of in-plane magnetic field, in which case the 
spin-quantisation axis lies parallel to the Q2DEG plane ($xy$-plane). 

The value of electric field at the surface is found from the Gauss theorem as
\begin{equation}
{\cal E}(z=0)=\frac{4\pi e}{\epsilon}(n+N_A z_d),
\label{eq:Ez0}
\end{equation}
and integrating Eq. (\ref{eq:Poi}) twice with the boundary 
conditions (\ref{eq:phigate}),(\ref{eq:Ez0})
yields the electrostatic potential,
\begin{eqnarray}
\!\!\!\!\!\!\!\phi(z)=\frac{1}{e}(E_c-E_{cs})-\frac{4\pi e}{\epsilon}(n+N_A z_d) z +
\frac{2 \pi e}{\epsilon}z^2 N_A + \nonumber \\ 
+\frac{4\pi e}{\epsilon}\int_0^z dz' \int_0^{z'} \left[\sum_{l,a,\alpha}n_{l,a,\alpha} \psi^2_{l,a,\alpha}(z'')
\right] dz''\,.
\label{eq:phi}
\end{eqnarray}
In turn, $\phi(z)$ enters the mean-field one-dimensional Hamiltonian which 
determines the carrier
motion in the directions perpendicular to the plane. At smaller densities,
only the two valleys with larger mass $m_\parallel$ corresponding to the 
$z$-axis motion are relevant
(ladder number $l=0$, valley degeneracy $\gamma_0=2$), with the corresponding
Hamiltonian 
\begin{eqnarray}
{\cal H}_{0,\alpha}&=&E_c- \frac{\hbar^2}{2m_\parallel}\frac{\partial^2}{\partial z^2}- 
e \phi(z)+
U\sum_{l,a}n_{l,a,-\alpha}\psi^2_{l,a,-\alpha}(z)
\nonumber \\
&&
-\frac{1}{2}H\sigma^z_{\alpha \alpha}\,, 
\label{eq:Ham}
\end{eqnarray}
Here, $H$ is the applied magnetic field in units of $g \mu_B$ (bare
$g$-factor times Bohr magnetone), and $\sigma^z$ is the Pauli
matrix. Owing to the finite thickness of Q2DEG, even an in-plane magnetic
field  leads to some orbital effects, as discussed
elsewhere\cite{Tutuc03,DasSarma2000} (experimentally, orbital effects of 
the in-plane field are indeed seen in magnetotransport measurements at 
small densities\cite{epl12}). These are expected to be minor and are omitted
in the present treatment. 

Eq. (\ref{eq:Ham}) includes the effects of short-range electron-electron 
interaction, Eq. (\ref{eq:contact3D}).
Presently,  considerable research effort is directed at exploring the 
possibility of manipulating valley polarisation 
(``valleytronics''\cite{Renard14}). Here, however, we are concerned with the 
spin 
degree  of freedom and for simplicity omit both the  repulsion between 
same-spin electrons from different valleys, and the dependence of $U$ on the 
valley indices. Yet we note that our approach can
be easily generalised to include these effects.
 
As the densities increase, electrons begin to populate also the four valleys 
($l=1$, $\gamma_1=4$) where
the larger mass $m_\parallel$ corresponds to an in-plane 
direction of motion\footnote{The role of the $l=1$ valleys was overlooked 
in Ref. \cite{jmmm}.}. These are 
described by the Hamiltonian ${\cal H}_{1,\alpha}$ which is given by Eq.  
(\ref{eq:Ham}) with the 
substitution $m_\bot \rightarrow m_\parallel$ on the r.\ h.\ s. 
Since the electrostatic potential confines the electrons to the vicinity of
the surface, the relevant (low-energy) parts of the spectra of the Hamiltonians
${\cal H}_{l,\alpha}$ are discrete,
\begin{equation}
{\cal H}_{l,\alpha}\psi_{l,a,\alpha}=E_{l,a,\alpha}\psi_{l,a,\alpha} \,.
\label{eq:Schroed}
\end{equation}
At $M=0$ the levels are spin-degenerate ($E_{l,a,\uparrow}=E_{l,a,\downarrow}$),
and form the two sequences (termed ``ladders''in Ref. \cite{Stern72}) 
corresponding to $l=0$ and $l=1$.
A spin-up electron at a level  $l,a$ interacts with a 
spin-down electron with a level number $l'\!, a'$ via  a 2D contact repulsion,
\begin{eqnarray}
{\cal U}_{2D}^{l,a;l'\!,a'\!}&=&U_{2D}^{l,a;l'\!,a'\!} \delta (x-x')\delta (y-y')\,, \nonumber \\
\,\,\,\,U_{2D}^{l,a;l'\!,a'\!}&=&U\int_0^\infty \psi^2_{l,a,\uparrow}(z) 
\psi^2_{l'\!,a'\!,\downarrow}(z) dz
\,.
\label{eq:U2D}
\end{eqnarray} 

Note that while a similar integral with the same value of spin projections for 
both wave functions does not vanish, the same-spin electrons with different
level indices do not interact. This is consistent with the underlying 
interaction ${\cal U}_{3D}$ being a contact one, as the presence of two 
same-spin electrons at the same point is forbidden.

Within the mean-field scheme, both the Hartree field due to ${\cal U}_{3D}$ 
and the self-consistent 
potential $\phi$ depend solely on $z$, hence  the energy of a Q2DEG carrier 
is a sum of the 
corresponding eigenvalue $E_{a,\alpha}$ of the transverse-motion Hamiltonian 
(\ref{eq:Ham}) 
and the free-particle contribution of the in-plane motion. In making this statement,
we neglect the relativistic effects (spin-orbit coupling) which is justified not
just because these are relatively small, but particularly because we are ultimately 
interested in thermodynamic quantities (magnetisation and susceptibility) 
which involve integrals over all directions of the in-plane momentum.
The 2D carrier densities for given level and spin indices  are thus 
given by
\begin{eqnarray}
n_{l,a,\alpha} &=&-\gamma_l \nu_l E_{l,a,\alpha}  \theta (-E_{l,a,\alpha})\,,
\label{eq:nalpha}\\
\nu_0&=&
\frac{m_\bot}{2\pi \hbar^2}\,,\,\,\,\nu_1=
\frac{\sqrt{m_\bot m_\parallel}}{2\pi \hbar^2}\,,
\label{eq:nu0}
\end{eqnarray}
where $m_\bot$ is the smaller effective mass and
$\theta$ is the Heaviside function.

Throughout the relevant range of parameter values, the spread of the wave 
functions $\psi_{l,a,\alpha}(z)$ in the $z$ direction is several orders of 
magnitude smaller that the depletion layer width $z_d$. This means that the
average values of $z$ for spin-up and -down carriers,
\begin{equation}
z_\alpha = \sum_{l,a} n_{l,a,\alpha} \int \psi_{l,a,\alpha}^2(z)z dz/n_{\alpha}
\label{eq:zalpha}
\end{equation}
are much smaller than $z_d$. Re-writing the last term in Eq. (\ref{eq:phi}) as
\[
\frac{4\pi e}{\epsilon}\sum_{l,a,\alpha}n_{l,a,\alpha} \left[
z\int_0^z\psi^2_{l,a,\alpha}(z') dz'- \int_0^z z' \psi^2_{l,a,\alpha}(z') dz' 
\right]\,,
\]
we then find that the condition that $\psi_{l,a,\alpha}(z)$ decays before the
value of $z$ reaches $z_d$ 
translates
into a useful relationship,
\begin{equation}
\frac{2 \pi e^2}{\epsilon}N_A z_d^2 = E_c-E_{cs}- \frac{4 \pi e^2}{\epsilon}
(n_\uparrow z_\uparrow + n_\downarrow z_\downarrow)\,.
\label{eq:zd}
\end{equation}

\begin{table}
%\begin{center}
\caption{\label{tab:values} Typical values of system parameters and material 
properties as used in the calculations. $m_e$ is the free electron mass. We 
chose the value of $U$ corresponding to $U_{on-site} \sim 3$ eV.}
\begin{tabular} {p{4.5cm} |l||l}
\hline
Quantity & ~ &Value\\
\hline
SiO$_2$ layer thickness &$d_{ox}$ & 10$^{-5}$ cm\\
Energy gap in Si & $E_c$ & 1.12 eV\\
Transverse (larger) effective mass & $m_{\parallel}$ & 0.916 $m_e$ \\
In-plane (smaller) effective mass  & $m_\bot$ & 0.19$m_e$\\
Acceptor density &$N_A$& 10$^{15}$ cm$^{-3}$\\
Critical density of MIT & $n_c$& 7$\cdot$10$^{10}$ cm$^-2$\\
3D contact repulsion & $U$& 7.5$\cdot$10$^{-34}$ erg$\cdot$cm$^{3}$\\
Dielectric constant of bulk Si & $\epsilon$ & 11.9 \\
Dielectric constant of SiO$_2$ & $\epsilon_{ox}$& 3.9\\
\hline
\end{tabular}
%\caption{\label{tab:values} Typical values of system parameters and material 
%properties as used in the calculations. $m_e$ is the free electron mass. We 
%chose the value of $U$ corresponding to $U_{on-site} \sim 3$ eV.}
%\end{center}
\end{table}

In the following, we describe the results of calculations performed within
this mean-field scheme in different regimes. The parameter values used are
given in Table \ref{tab:values}.  In order to facilitate convergence of the numerical scheme, 
we made use of some of the algorithms employed previously in the 
non-interacting, 
zero-field case\cite{Stern70}. The $z$-coordinate is discretised, the system 
of Poisson and Schroedinger equations for $z>0$ is solved, and its solution is 
fed back 
into the Hamiltonian for the next iteration.  The infinite potential barrier at
$z \leq 0$ is modelled by cutting off the hopping to the $z=0$ point of the
discretised $z$-axis from the $z>0$ side.

 In addition to including the effects of
$U$ and $H$, an important difference from the previous calculations (including
Ref. \cite{Stern72}) is that instead of fixing $E_{cs}$, we set the 
problem in a more precise way, fixing $\phi_{gate}$ and solving for 
$E_{cs}$, $n$, $M$ and $z_d$. While the numerical calculations become 
more involved in this formulation of the problem\footnote{
 In practice, we first fix $E_{cs}$ and solve  
for $n$, $M$ and $z_d$, which can be done by feeding the results 
of each subsequent iteration back into the mean-field equations 
(\ref{eq:nalpha}) 
and (\ref{eq:zd}) (cf. Ref. \cite{Stern70}); the appropriate 
value of $E_{cs}$ is 
then found as the root of Eq. 
(\ref{eq:phigate}).},
%\cite{calcul}, 
it  corresponds to the actual 
measurement setup.
Physically, the difference becomes apparent in the
phenomenological treatment of the strongly-correlated case in 
Sect. \ref{sec:dolgo} (where the bandwidth, and hence $E_{cs}$, vary 
self-consistently), and also 
in the case of large magnetisation values encountered in Sects. \ref{sec:metal} 
and \ref{sec:field}.   

Indeed, if at a fixed value of $E_{cs}$ and at $U=0$ the field $H$ is 
increased beyond the
value $H^\downarrow_{l,a}$ corresponding to a full spin polarisation of carriers with a 
certain transverse-ladder and level indices $l,a$, the value of $n$ would 
begin to increase
as $\delta n = (H-H^\downarrow_{l,a})/(\gamma_l \nu_l)$ (the value of $\partial H/\partial n$ would be renormalised in an interacting system). This is unphysical 
as in reality 
this variation of $n$ is for the most part suppressed by the large capacitance
$C$ in Eq. (\ref{eq:phigate}). We find that the relative 
change in $n$ is in fact rather small (see
below, Sect. \ref{sec:field}). 

Thus, the appropriate mean-field thermodynamic potential, which is minimised
by the suitable mean-field solution, corresponds to  fixing $\phi_{gate}$,
rather than $n$:
\begin{eqnarray}
&&\!\!\!\!\!\!\!\!\!G=
\sum_{l,a,\alpha} n_{l,a,\alpha}\left\{ 
 E_{l,a,\alpha}+ \frac{n_{l,a,\alpha}}{2 \gamma_l\nu_l} + 
\frac{e}{2}\int_0^\infty\psi_{l,a,\alpha}^2(z)
  \phi(z)dz\right\}-\nonumber \\ 
&&-\sum_{l,l'\!,a,a'\!}\nonumber U_{2D}^{l,a;l'\!,a'\!}n_{l,a,\uparrow}n_{l'\!,a'\!,\downarrow}-
\frac{eN_A}{2}
  \int_0^{z_d} \phi(z)dz\\
&&-\frac{e}{2}(n+N_A z_d)\phi_{gate}\,. 
\label{eq:G}
\end{eqnarray} 
The three terms in the first line are the energies of $z$-axis and in-plane 
motion of the Q2DEG carriers, and the correction to exclude the double-counting
of their electrostatic energy. Double-counting of the interaction energy is
corrected by the first term in the second line, whereas the second term is
the electrostatic energy of immobile electrons in the depletion layer. The last
term, $-Q_{gate} \phi_{gate}/2$, corresponds to our choosing $\phi_{gate}$ as an 
external variable. 

%It should be noted that when calculating the inversion-layer carrier energies we
%did not include the electrostatic image potential\cite{Ando82}. This is 
%because of
%the metallic behavior\cite{Kravchenko94} of Si-MOSFETs in the range of density values which is of
%interest to us here. From the point of view of electrostatics, a delocalised 
%electron is not a point charge, but rather a uniformly charged 
%layer\cite{linear}. This makes
%the electrostatic problem perfectly one-dimensional, and image charges therefore
%do not arise. While leaving this issue for future investigation, we note that 
%this effect {\it lowers the energy of metallic carriers} relative to that of 
%localised ones.

A discussion of the applicability of our mean field scheme as outlined above 
is relegated to the Appendix. We will now turn to the results obtained 
in different regimes.

\section{Electrical quantum limit: the single-level solution}
\label{sec:single}

If the value of the gate voltage $\phi_{gate}$ is not too large, only the lowest
quantum level $E_{0,0,\alpha}$ of the $z$-axis motion for each spin direction 
can lie below the 
chemical potential and be populated by the Q2DEG carriers:
\begin{eqnarray}
n_{\alpha} &=&-2 \nu_0 E_{0,0,\alpha}  \theta (-E_{0,0,\alpha})\,,\,\,\,
\label{eq:nalpha0} \\
n_{l,a,\alpha}&=&0 \,\,{\rm for}\,\, a \geq 1\,\, {\rm or}\,\, l \neq 0.
\nonumber 
\end{eqnarray}
This situation, which is termed electrical quantum limit, is somewhat simpler 
to analyse than the full multi-level case, and we will consider it first
in order to illustrate certain key features of our mean-field results and 
underlying physical mechanisms. Moreover, we find it expedient to formally 
allow for values of $\phi_{gate}$ (or, equivalently, of $n$) to increase 
beyond the range where the electrical quantum 
limit is realised (the latter corresponds to lower carrier densities, 
$n \leq 3.2\cdot 10^{12}$ cm$^{-2}$, see Sect. \ref{sec:metal}). This is 
accomplished by using Eq.  
(\ref{eq:nalpha0}) in place of
Eq. (\ref{eq:nalpha}), while keeping the rest of the mean field scheme intact.
For quantitative results in the larger-density case of 
$n > 3.2\cdot 10^{12}$ cm$^{-2}$, the reader should refer to 
Sect. \ref{sec:metal} below. Since within the present section the ladder and 
level indices
of all quantities are always equal to zero, we will suppress these.

Let us first briefly recall the usual Stoner picture, applicable both in the 
three-dimensional bulk and in the case of a perfectly 2D carriers. The latter
 have no $z$-axis
degree of freedom and  interact via contact potential, ${\cal U}_{2D}=U_{2D} 
\delta (x-x')\delta (y-y')$. At the mean-field level, the effect of interaction
is additive with that of the applied field $H$, increasing the energy shifts
of the two spin subbands (Zeeman splitting). The wave functions (which in the 
2D case are given by products of $\delta(z)$ and the in-plane Bloch wave) are 
unaffected, and  one readily finds the magnetic susceptibility, which in the
2D case is given by
\begin{equation}
\chi_0 = 
\frac{\nu_0}{1-2\nu_0U_{2D}}\,.
\label{eq:2dstoner}
\end{equation}
As long as both the 2D density of states $\nu_0$ and $U_{2D}$ remain constant, 
$\chi_0$ does not depend on density. If either of these can be varied to the 
extent that the denominator of Eq. (\ref{eq:2dstoner}) vanishes, the ensuing
divergence of $\chi_0$ suggests a ferromagnetic transition. Owing to the 
independence of $\nu_0$ on the carrier energy in the 2D case, this critical 
point has a peculiar character of a discontinuous transition with no 
hysteresis. Specifically, everywhere in the 
ferromagnetic phase the mean-field free energy minimum is attained in the 
fully spin-polarised state, whereas at the transition point itself the free
energy does not depend on the magnetisation. Thus the magnetisation shows a 
jump at the transition point, simultaneous with vanishing of both the spin 
stiffness (from the ferromagnetic side) and inverse susceptibility.

These properties are strongly modified in the case of the Q2DEG as found in a
Si inversion layer. First, note that the interaction strength $U_{2D}$ is 
given by Eq. (\ref{eq:U2D}) and depends on the density $n$. This is 
illustrated by the variational treatment, where the solution to Eq. 
(\ref{eq:Schroed}) is sought in the form of an {\it ansatz}, 
\cite{Stern72,Fang1966}
\begin{equation}
\psi_{var}(z)=\sqrt{\frac{b^3}{2}}z \exp(-bz/2)\,,
\label{eq:Stern}
\end{equation}
yielding $U_{2D}^{var}=3 U b /16$. The value of $b$ is chosen by minimising the
thermodynamic potential, Eq. (\ref{eq:G}), which yields
\begin{equation}
\frac{\hbar^2 b^3}{4 m_\parallel} +\frac{3}{64}Un b^2- \frac{12 \pi e^2}{\epsilon} \left(N_A z_d + \frac{11}{32}n \right) =0\,.
\label{eq:Sternb}
\end{equation}

The difference from the $U=0$ result of Ref. \cite{Stern72} is in
the addition of the second term on the l.\ h.\ s. This results in a
slight decrease of the value
of $b$ (and hence in an increase of $z_0 \equiv z_\uparrow=z_\downarrow = 3/b$) in
comparison to the non-interacting case. Similar to the $U=0$ case, we find
that the variational value of $E_0=E_\uparrow=E_\downarrow$,
\begin{equation}
E_0=E_{cs}+ \frac{\hbar^2 b^2}{8 m_\parallel}+ 
\left(N_a z_d +\frac{11}{16}n \right)\frac{12 \pi e^2}{\epsilon b} + 
\frac{3}{32} U n b\,,
\label{eq:E0var}
\end{equation}
 closely approximates
the numerical result.

In both $U=0$ and $U>0$ cases, the value of $b$ increases with increasing $n$. 
This is due to the increase of  the ratio $n/N_Az_d$ 
($z_d$ only weakly depends on $n$), which leads to a progressively 
larger part of
the electrostatic field of the gate being screened by the mobile carriers 
{\it within} the layer of the 
Q2DEG (and not elsewhere within the depletion layer). Hence the potential
$\phi(z)$ becomes steeper at small $z$, resulting in smaller $z_0$ and 
larger $b$ and $U_{2D}$. 

\begin{figure}  \sidecaption
\includegraphics[width=.49\textwidth]{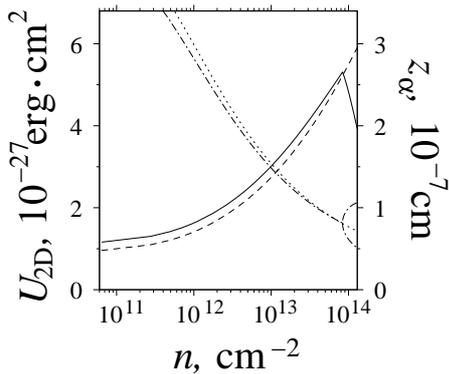}
\caption{\label{fig:U2D} Density dependence of the effective 2D short-range
repulsion $U_{2D}$ (solid line, left scale) and of the average carrier distance
from the surface $z_0=z_\uparrow=z_\downarrow$ 
(dashed-dotted line, right scale), calculated at $H=0$ 
by solving the mean field equations numerically. Above the critical density
$n_1\approx 8.4 \cdot 10^{13}$ cm$^{-2}$, the dashed-dotted line splits into two,
corresponding to $z_\downarrow>z_\uparrow$.
 Dashed and dotted lines show
the values of $U_{2D}$ and $z_0$, respectively, obtained using the variational
wave function $\psi_{var}(z)$, Eq. (\ref{eq:Stern}).
}
\end{figure}

The decrease of $z_\alpha$ and increase of $U_{2D}$ are also found in the 
numerical solution of the mean-field equations in the paramagnetic state,
{\it i.\ e.}, below the critical density $n_1\approx 8.4 \cdot 10^{13}$ 
cm$^{-2}$.  The dependence of $U_{2D}$ and 
$z_\alpha$ on $n$ is depicted in Fig. 
\ref{fig:U2D},
showing both variational and numerical results. We thus conclude that owing
to the Q2DEG layer thinning, the quantity $\chi_0$ in the paramagnetic phase
must be increasing with 
$n$, as indeed seen in numerical and variational results (see below, Figs. \ref{fig:chivar} and \ref{fig:susce}).

\begin{figure}  \sidecaption
\includegraphics[width=.49\textwidth]{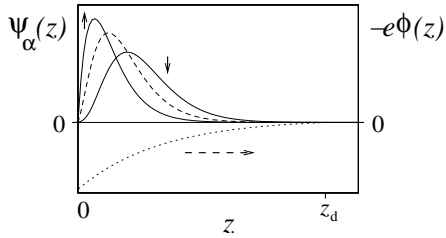}
\caption{\label{fig:scheme} Interaction-induced evolution of 
transverse-motion wave functions with increasing magnetisation $M$. 
The  spin-up and -down wavefunctions at $M=0$ coincide (unpolarised case,
shown schematically by the dashed line). In the presence of interaction $U$, 
they split at $M>0$ 
(solid lines). The electrostatic potential energy, $-e\phi(z)$, is 
shown for simplicity as a single dotted line [in reality, an increase of 
$M$ is accompanied 
by a small self-consistent change in $\phi(z)$ and in the value of $z_d$]. }  
\end{figure}

An additional effect arises in the interacting case, $U>0$, with the result 
that the actual magnetic susceptibility $\chi=\partial M/\partial H$ is no 
longer given by Eq. (\ref{eq:2dstoner}). Indeed, it is easy to see that at 
$U>0$,
the appearance of a  spin 
polarisation must be accompanied by a change in
the transverse wave functions $\psi_\alpha(z)$  -- a phenomenon which does 
not occur in the 
familiar Stoner picture as outlined above. At the level of our mean-field 
Hamiltonian,  Eq. (\ref{eq:Ham}), the effect of interaction $U$ is that an 
electron feels an additional potential bump 
[the fourth term in Eq.  (\ref{eq:Ham})], centred around the peak of the 
opposite-spin wave function. In the absence of polarisation ($M=0$), these 
peaks are located roughly at $z \sim z_0=z_\uparrow=z_\downarrow$ and are 
identical for spin-up and -down carriers. When the magnetisation 
differs from zero (either spontaneously in the ferromagnetic phase or due to 
an applied magnetic field), the height (proportional to $n_\uparrow$) of the 
potential bump in the Hamiltonian ${\cal H}_\downarrow$ of the spin-minority 
electrons increases, pushing these further away from $z_\uparrow$ in the 
direction of larger $z$ and increasing $z_\downarrow$ 
(note that now $z_\downarrow> z_\uparrow$, as seen in Fig. \ref{fig:U2D} for 
$n>n_1$ ). The associated bump in 
${\cal H}_\uparrow$ (although somewhat reduced in size, due to a reduction of 
$n_\downarrow$) is no longer centred
at the peak of spin-majority electrons distribution -- rather, it is ``pushing''
on these electrons from the side of larger $z$, 
leading to a reduction of $z_\uparrow$.  This situation, which is shown 
schematically in Fig. \ref{fig:scheme}, 
leads to  decreasing the overlap 
between spin-up and spin-down wave-functions, hence to decreasing $U_{2D}$
[see Eq. (\ref{eq:U2D})]. This 
behaviour of the numerical 
result for $U_{2D}$
is clearly reflected in Fig. \ref{fig:U2D} for $n>n_1$ (where $M>0$, see 
the inset in Fig. \ref{fig:susce} below). 
Ultimately the value of thermodynamic potential $G$
(Eq. (\ref{eq:G})) is reduced in comparison with the case where no allowance 
is made for the change of $\psi_\alpha(z)$ with $M$. In other words, as a 
result of wave functions profile change it costs less energy to form a 
non-zero magnetisation, which translates into an increased value of 
susceptibility $\chi$ and into a decreased critical value of the interaction 
$U_{2D}(n)$ (evaluated at $M=0$), required to destabilise the paramagnetic 
state\footnote{
Note that this  wave functions change is not restricted to electrons in the 
vicinity of the 2D Fermi surface. This implies that Fermi liquid theory cannot
be used to evaluate magnetic susceptibility, and the conventional 
Fermi-liquid expression for $\chi$, 
which can be 
viewed as an analogue of Eq. (\ref{eq:2dstoner}), is inapplicable 
in this case.}.
In the purely 2D case, the latter is determined by a condition [cf. 
Eq.(\ref{eq:2dstoner})]
\begin{equation}
1-2\nu_0 U_{2D}  =0\,,
\label{eq:oldstoner}
\end{equation}  
known as the Stoner criterion. As we already mentioned, what is varied 
in the actual measurements is the gate voltage $\phi_{gate}$, which in turn
causes the variation of $n$, directly accessible by measuring the Hall voltage. 
Hence the relevant quantity is the value of $n$, corresponding to the 
ferromagnetic transition. Owing to the dependence of $U_{2D}$ on $n$, 
the l.\ h.\ s. of Eq. (\ref{eq:oldstoner}) for a given $U$ may vanish at a 
certain critical value
density, $n_0$ (which is either very large or even infinite for our values
of parameters).
 In reality, we find that the Stoner criterion is relaxed, 
{\it viz.,}
the l.\ h.\ s. of Eq. (\ref{eq:oldstoner}), 
is still positive at the critical 
density, $n=n_1$. This is due to  self-consistent dependence of the
transverse-motion wave functions on  magnetisation 
$M$, as discussed above.

These ideas can be illustrated with the help of variational wave functions.
As explained above, using the wave function (\ref{eq:Stern}) leads to
\begin{equation}
\chi_0^{var} = 
\frac{\nu_0}{1-3 U b(n) \nu_0 /8}\,,\,\,\,\,
1-3 U b(n_0^{var}) \nu_0 /8=0\,,
\label{eq:chi0var}
\end{equation} 
where $n_0^{var}$ is the corresponding critical density. Let us now allow the
spin-up and -down wave functions to differ form each other at $M\neq 0$, by 
writing, instead of
Eq. (\ref{eq:Stern}),
\begin{equation}
\psi_\alpha(z)=\sqrt{\frac{b^3_\alpha}{2}}z \exp(-b_\alpha z/2)\,.
\label{eq:sternalpha}
\end{equation}
Here, we are interested in the limit of small polarisation, $M \ll n$. 
Thus,  $b_{\uparrow,\downarrow}-b=\pm b_1$ is a small spin-dependent correction
to the value of $b$ which  solves Eq. (\ref{eq:Sternb}) at $M=0$ and $H=0$. 
We then substitute Eq. (\ref{eq:sternalpha}) into  
Eq. (\ref{eq:G}), which includes re-calculating the variational energies 
$E_{\alpha}=\int_0^\infty \psi_\alpha(z) {\cal H}_\alpha \psi_\alpha(z) dz$.
To leading-order in $b_1$, $M$, and $H$, the thermodynamic potential $G$ 
acquires a correction,
\begin{eqnarray}
&&\delta G=\left\{\frac{\hbar^2n}{8m_\parallel}+ \left(12N_A z_d +\frac{21}{16}n 
\right)\frac{\pi e^2n}{\epsilon b^3}-\frac{9 U n^2}{64 b} +
\right. \nonumber \\
&& \left. +\frac{9}{2048} \nu_0U^2 n^2 \right\} b_1^2 - \left(1-\frac{3}{8} \nu_0 b U \right) \left\{ \frac{3}{32}Un M b_1 + 
\right. \nonumber \\
&& \left.
+\frac{3}{16}U b M^2 +MH \right\}+ \frac{3}{32}U n \nu_0 b_1 H + 
\frac{1}{2} \nu_0 H^2\,.
\label{eq:deltaG}
\end{eqnarray}
Note that $b_1$, $H$, and the magnetisation $M=(n_\uparrow-n_\downarrow)/2$ are 
not mutually independent. Indeed, $M$ is obviously
determined by the first-order correction to the variational energy $E_0$,
\begin{eqnarray}
M&=&-\nu_0 (E_\uparrow-E_\downarrow)=\nu_0 H + \frac{3}{8} \nu_0 U b M -
\nonumber \\
&& - \left\{ \frac{\hbar^2 b}{4 m_\parallel} -\left( N_A z_d + \frac{11}{32}n 
\right)\frac{12 \pi e^2}{\epsilon b^2} \right\}2 \nu b_1\,,
\end{eqnarray}
or, with the help of Eq. (\ref{eq:Sternb}),
\begin{equation}
\left(1-\frac{3}{8} \nu_0 b U \right)M-\left(\frac{3}{32} Un\nu_0 b + \nu_0 H 
\right)=0\,.
\end{equation}
We can now use this to exclude $b_1$ in Eq. (\ref{eq:deltaG}). 
Minimising $\delta G$ with respect to $M$ then yields $M= \chi^{var} H$,
with the corresponding susceptibility
\begin{eqnarray}
\chi^{var}&=&\frac{\nu_0}{1-3 U b(n) \nu_0 /8-L}\,,
\label{eq:chivar}
\\
L&=&\frac{9 \nu_0 U^2 n b^2}{2048}\left\{\frac{\hbar^2 b^2 }{8m_\parallel}+
\right. \nonumber \\
&& \left. +\left(N_az_d+ \frac{7}{64}n \right)\frac{12 \pi e^2 }{\epsilon b}-
\frac{9}{64}Unb \right\}^{-1}\,.
\end{eqnarray}
We see that the effect of wave function changing with $M$ gives rise to the
last term in the denominator in Eq. (\ref{eq:chivar}) 
[cf. Eq. (\ref{eq:chi0var})], and therefore leads to the susceptibility 
increase. The second term in the denominator equals $2\nu_0 U_{2D}^{var}$, and
the ratio $L/(2\nu_0 U_{2D}^{var})$ is roughly of the order of 
$nU_{2D}/(E_0-E_{cs})$.
Here, $nU_{2D}/2$ is the net scale of the energy of the contact interaction, 
whereas $E_0-E_{cs}$ is the energy of quantised motion along the $z$-axis 
[see Eq.(\ref{eq:E0var})].    

These variational results are illustrated in Fig. \ref{fig:chivar}, where the 
solid line represents
Eq. (\ref{eq:chi0var}), which uses the {\it ansatz} (\ref{eq:Stern}) for the 
wave function and does not
allow for a wave function change with increasing $M$. The variational 
susceptibility $\chi_0^{var}$ slowly 
increases with $n$ from the non-interacting value of $\chi_0^{var} = \nu_0$, 
reflecting the increase of 
$U_{2D}$ as discussed above. Within a very broad range of $n$, it does not 
show any
critical behaviour: indeed, at $n$ as large as $3.5 \cdot 10^{14}$ cm$^-2$, 
$\chi_0^{var}/\nu_0$ reaches the value 
of only 1.6. On the other hand, the quantity $\chi^{var}$ [dashed line; see Eq. 
(\ref{eq:chivar})] deviates upwards from 
$\chi_0^{var}$ and diverges at $n_{var} \approx 1.13 \cdot 10^{14}$  
cm$^-2$, suggesting a ferromagnetic 
transition. This is a consequence of the polarisation dependence of 
the wavefunctions (\ref{eq:sternalpha}),
as outlined above. 
\begin{figure}  \sidecaption
\includegraphics[width=.49\textwidth]{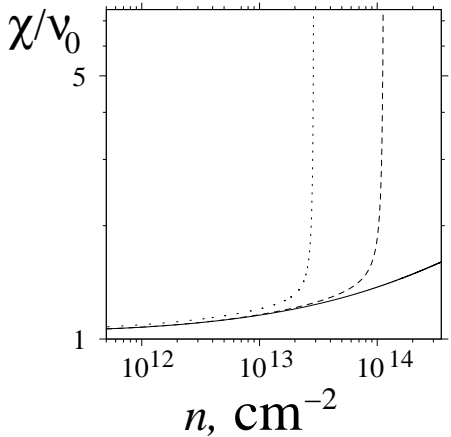}
\caption{\label{fig:chivar} Variational results for the density dependence of 
normalised susceptibility, $\chi/\nu_0$, in the  single-level case.
Solid line corresponds to Eq. (\ref{eq:chi0var}) and does not include the 
effect of $H$ on the wave function shape. Dashed line represents Eq. 
(\ref{eq:chivar}), obtained using the field-dependent variational wave 
functions (\ref{eq:sternalpha}). Dotted line corresponds to an improved 
{\it ansatz}, Eq. (\ref{eq:wf160}). 
}
\end{figure}
We note that the difference between $\chi^{var}$ and $\chi_0^{var}$ becomes 
appreciable only at large densities $n$, and the critical value $n_{var}$
is also very large. This is due to our chosen wave functions shape, Eq. 
(\ref{eq:sternalpha}). Indeed, it is clear that the way these wave functions
are changed with $H$ is far from optimal. Much lower result for the critical 
density ($\tilde{n}_{var} \approx 2.89 \cdot 10^{13}$  cm$^{-2}$) is obtained when using an {\it ansatz}
which includes additional parameters $\kappa_{\uparrow,\downarrow}$:
\begin{equation}
\tilde{\psi}_\alpha(z)\propto z \sqrt{1+ \kappa_\alpha b_\alpha^2 z^2} \exp(-b_\alpha z/2)\,.
\label{eq:wf160}
\end{equation}
This results in a somewhat cumbersome expression for susceptibility, which
is given in Ref. \cite{jmmm}; in Fig. \ref{fig:chivar}, the corresponding
value is plotted with a dotted line. It does not merge
 with $\chi_0^{var}$ even at low densities 
because the optimal value of coefficient 
$\kappa_\uparrow=\kappa_\downarrow$ at $M=0$ differs from 
zero\cite{jmmm}.

The numerical solution of the mean-field equations in the single-level case 
yields the solid
line in Fig. \ref{fig:susce} (for comparison, the dashed line shows the 
value of  $\chi^{var}$). The numerical
result shows  critical behaviour, with the corresponding critical $n_1$ 
in the interval between 
$\tilde{n}_{var}$ and ${n}_{var}$. Thus, we conclude that the latter 
two variational approximations
respectively overestimate and underestimate the ferromagnetic tendencies. 
The importance of the wave-function
change with $M$ in case of numerical results is illustrated by the 
dotted line in Fig. \ref{fig:susce},
which shows the value of $\chi_0$, Eq. (\ref{eq:2dstoner}), 
computed using the numerically calculated
value of $U_{2D}$ at $M=0$[see Eq. (\ref{eq:U2D}) and Fig. \ref{fig:U2D}]. 
In other words, when calculating
$\chi_0$  we used the exact mean field wave functions for $M=0$. 
Thus, the quantity $\chi_0$ is defined only at 
$n<n_1$, and we see that it remains smaller than the actual 
susceptibility $\chi$ and does not show any
tendency toward criticality (similarly to $\chi_0^{var}$ 
in Fig. \ref{fig:chivar}). The interaction energy per a  
Q2DEG carrier can be estimated as $n U_{2D}/2$ 
[see Eq. (\ref{eq:Ham})]
and decreases with $n$. Hence at small $n$ it eventually becomes much 
smaller than the energy $E_0-E_{cs}$ of 
the transverse motion, measured from the bottom of the potential well 
[estimated as $\hbar^2/(m_\parallel z_0^2)$]. 
In this situation, transverse carrier motion is no longer affected by 
the interaction $U$, and in particular a 
change in $M$ does not lead to an appreciable change of transverse wave 
functions. Indeed, we see that  
$\chi$ and $\chi_0$ become almost undistinguishable at densities 
below $\sim 5 \cdot 10^{11}$ cm$^-2$.  

\begin{figure}  \sidecaption
\includegraphics[width=.49\textwidth]{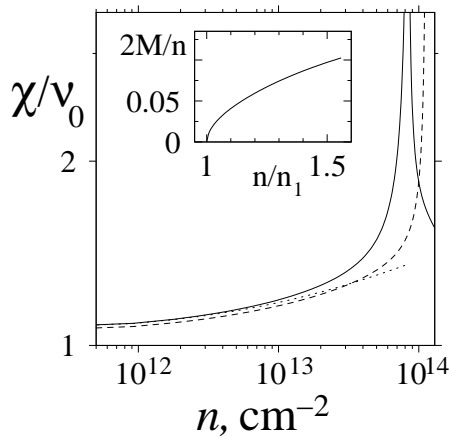}
\caption{\label{fig:susce} Numerical results for the density dependence of 
normalised susceptibility, $\chi/\nu_0$, in the  single-level case at $H=0$ 
(solid line).
We also show the susceptibility values corresponding to 
Eq. (\ref{eq:chivar}) (dashed line) and
to Eq. (\ref{eq:2dstoner}), using the numerical result for 
$U_{2D}$ (dotted line).
The inset shows the numerical result for the degree of spin 
polarisation, $2M/n$,
which arises above the ferromagnetic transition at 
$n_1 \approx  8.4 \cdot 10^{13}$ cm$^{-2}$.}
\end{figure}

The density dependence of magnetisation $M$ in the ferromagnetic phase at
 $n>n_1$ is shown in the inset of
Fig. \ref{fig:susce}. It looks reminiscent of a typical mean field behaviour 
of an order parameter, yet as
explained above this is {\it not} what is found in the Stoner treatment of
 a purely 2D case, where a jump in
$M$ is obtained. The difference is due to the transverse wave functions 
changing with increasing $M$: the resultant decrease of $U_{2D}$ moderates
the increase of $M$ with density.

In the preliminary publication \cite{jmmm}, we used a larger value of $U$
(4/3 of the value used presently), leading to smaller values of numerical and
variational critical densities. We find that the result of Ref. \cite{jmmm}
for the numerical solution equals 0.64   of our present $n_1$, and similarly 
for the variational ansatz, Eq. (\ref{eq:wf160}), Ref. \cite{jmmm} yields 
the critical density
of  $0.66 \tilde{n}_{var}$. We conclude that critical density is strongly
dependent on $U$.

As already mentioned, the simplified treatment described in this section, while illuminating, does not
apply in the two important limiting cases, {\it viz.}, the metallic behaviour at high densities and the
correlated regime immediately above the MIT. We will now consider these in more detail.

%\section{Phenomenological treatment in the vicinity of the MIT}
\section{Low carrier densities above the MIT}
\label{sec:dolgo}

Within the simplified single-level treatment of Sect. \ref{sec:single},
the obtained value of magnetic susceptibility was found to increase with
increasing density $n$, eventually reaching a ferromagnetic instability
deep in the high-density region 
(see Figs. \ref{fig:chivar} and \ref{fig:susce}). We note that at the low 
densities above the MIT, only the lowest transverse-motion level lies below
the chemical potential, hence Eq. (\ref{eq:nalpha0}), used in  
Sect. \ref{sec:single}, is certainly valid. In this low-density range, the 
computed value of susceptibility as plotted in Fig. \ref{fig:susce} 
(solid line) only slightly deviates from the non-interacting result, 
$\chi \equiv \nu_0$ (see the dashed-dotted line in
Fig.  \ref{fig:dolgo} below). 
However, the approach used in Sect. \ref{sec:single} is contingent 
upon the validity of 
%our mean-field scheme, {\it viz.}, upon 
the assumption that after taking into account both
Coulomb and contact interactions {\it on average}, the in-plane carrier motion
can be treated as free. The latter
% Such an assumption regarding the Coulomb interaction
becomes invalid at low densities, where the dimensionless parameter 
$r_s=m_\bot e^2/(\epsilon \hbar^2 \sqrt{\pi n})$ (relative strength of the long-range Coulomb interaction) significantly exceeds 1. 
Since at $n \approx n_c= 7 \cdot 10^{10}$ 
cm$^{-2}$
(see Table \ref{tab:values}) we find $r_s \approx 6.4$, our mean field scheme 
as outlined in Sect. 
\ref{sec:mf} is indeed inapplicable in this region. Here, we wish to argue 
that a 
phenomenologically-motivated modification should be introduced in the 
self-consistent mean field scheme in this regime.

Recently, it has been noted\cite{Dolgopolov15} that the available data for
the effective mass, susceptibility, and saturation field value in Si-MOSFETs
above the MIT
can be described
phenomenologically by a 2D non-interacting Fermi gas with a renormalised
in-plane mass:
\begin{equation}
\tilde{m}_\bot = m_\bot \frac{n}{n-n_c}\,.
\label{eq:dolgo}
\end{equation}
This behaviour was anticipated theoretically\cite{Dolgopolov02}, 
and discussed in the general 
context of metal-insulator transitions\cite{Dobro12}. Similar results were also
obtained by
radiative spectra measurements on GaAs/AlGaAs 
heterostructures\cite{Kukushkin15}. In addition, higher-temperature entropy 
measurements\cite{Kuntsevich15} on a Si-MOSFET sample yield an effective mass
peak at low densities. The peak becomes more pronounced when the temperature is
lowered, and this effective mass enhancement is in a qualitative agreement 
with the low-temperature results as 
described by Eq. (\ref{eq:dolgo}). %Eq. (\ref{eq:dolgo}) 
The latter equation leads to a renormalisation
of the density of states,
\begin{equation}
\tilde{\nu} \equiv \frac{\tilde{m}_\bot}{2\pi \hbar^2}=\frac{n}{n-n_c} \nu_0\,
\label{eq:nutilde}
\end{equation}
and (in the absence of the short-range interaction $U$) to the Pauli in-plane
susceptibility\cite{Dolgopolov15},
\begin{equation}
\chi_{P}=\tilde{\nu}\,, 
\label{eq:Pauli}
\end{equation}
which diverges at the MIT (at $n=n_c$). The latter is due to the 
effective band narrowing, and does not necessarily imply a magnetic 
instability (in agreement also with Ref. \cite{Finkelstein05}).

In the low-density region of $n \stackrel{<}{\sim} 10^{11}$ cm$^{-2}$, 
the average
distance between carriers is large in comparison with the inversion layer 
thickness (of the order of $10^{-6}$ cm). It is then natural to expect that
while the long-range correlations are in fact prominent (as indicated by large values
of $r_s$), they affect the in-plane motion of the carriers only, whereas
 the finite carrier motion along the $z$-axis is still determined by a nearly 
triangular self-consistent potential $\phi(z)$. Hence it appears that the effects of an additional short-range
interaction $U$ can be probed within the Hartree scheme as before.
%(and with greater accuracy, as the relative strength of short-range interaction
%decreases with decreasing density). 
The only modification which needs to 
be 
introduced in the mean-field scheme of Sect. \ref{sec:mf} is the substitution
of $\tilde{\nu}$ in place of $\nu_0$ in Eq. (\ref{eq:nalpha}) 
[or equivalently in Eq.  (\ref{eq:nalpha0})]. We emphasise that this approach does not constitute a self-contained
theoretical treatment (hitherto missing), which should include both interactions from the start.
In reality, what we attempt here is  a phenomenological estimate, whose results underline the necessity of constructing
a proper theoretical description. 

When neglecting the wave function dependence on magnetisation (which
is indeed justified in this regime, see below), we obtain instead of Eq.
(\ref{eq:2dstoner}):
\begin{equation}
\chi_0 = 
\frac{\tilde{\nu}}{1-2\tilde{\nu}U_{2D}}\,.
\label{eq:2dstonertilde}
\end{equation}
As the density is lowered toward $n_c$, the value of $U_{2D}$ stays finite 
while $\tilde{\nu}$ diverges, signalling a ferromagnetic instability at
\begin{equation}
n_*=n_c \left(1+2 \nu_0U_{2D} \right)\,.
\end{equation} 
In order to roughly estimate the difference between this transition and the MIT,
one can again use the variational {\it ansatz} (\ref{eq:Stern}), which yields
 $U_{2D}^{var}=3 U b /16$. The second term in Eq. (\ref{eq:Sternb}) for the 
variational parameter $b$ is now negligible, whereas in other terms $n$ and 
$N_Az_d$ are of
the same order of magnitude as $n_c$. Omitting all factors of order of unity, we obtain an order-of-magnitude estimate,
\begin{equation}
n_*-n_c \sim U\nu_0 n_c^{3/2}r_s^{1/3}\,.
\label{eq:nstar}
\end{equation}

Variational and numerical results for susceptibility are shown in Fig. 
\ref{fig:dolgo}. As explained above, at $U=0$ the (Pauli) susceptibility
$\chi_P$, Eq. (\ref{eq:Pauli}), diverges at $n=n_c$ but does not show any 
ferromagnetic singularity 
at $n>n_c$ (dashed line
in Fig. \ref{fig:dolgo}). Numerical solution of the mean-field equations
[with renormalised density of states $\tilde{\nu}$, 
see Eq. (\ref{eq:nutilde})] yields the value 
of $\chi$ showed by the solid line, with a ferromagnetic
instability at $n_* \approx 7.43 \cdot 10^{10}$ cm$^{-2}$. Hence taking into 
account the short-range $U$ brings about the ferromagnetic transition above the
MIT. The dashed-dotted line shows the results obtained within the approach of
Sect. \ref{sec:single} with the same value of $U$ but without renormalising
the density of states (i.\ e., using $\nu_0$ rather than $\tilde{\nu}$).
While the dashed-dotted and solid lines eventually merge at higher
$n$ (where the effects of long-range correlations are weak), the dashed-dotted
line remains featureless all the way down to $n=n_c$.

Similar to Sect. \ref{sec:single} above, a comparison with the results of Ref.
\cite{jmmm} allows to verify the dependence of $n_*$ on $U$. We find
that the result of Ref.
\cite{jmmm}  for $(n_*-n_c)/n_c$ is about 1.3 times larger than the one 
obtained herein, roughly agreeing with Eq. (\ref{eq:nstar}). 

\begin{figure}  \sidecaption
\includegraphics[width=.49\textwidth]{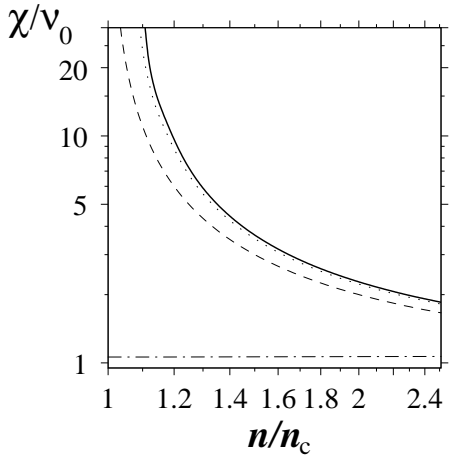}
\caption{\label{fig:dolgo} Magnetic susceptibility $\chi$ in the units of
  the bare density of states $\nu_0$ [see Eq. (\ref{eq:nu0})]  in the 
 low-density region above the MIT. In-plane carrier mass is renormalised 
according to 
  Eq.(\ref{eq:dolgo}). Solid, dashed, and
  dotted line correspond, respectively, to the numerical solution of the 
mean-field  equations, Pauli susceptibility  (\ref{eq:Pauli}), and
 the variational result using 
Eqs.(\ref{eq:Stern}),(\ref{eq:2dstoner}) with the substitution
$\nu_0 \rightarrow \tilde{\nu}$. 
Dashed-dotted line shows the numerical solution of the mean-field equations
with the unrenormalised density of states $\nu_0$.} 
\end{figure}

We note that on 
the scale of the plot, the
numerical value of $\chi(n)$ (solid line in Fig. \ref{fig:dolgo}) is indistinguishable from $\chi_0$, 
Eq. (\ref{eq:2dstonertilde}). This is because in the low-density regime,
the characteristic energy scale $nU_{2D}$ of the short-range interaction is
much smaller than the ground-state energy $E_0-E_{cs}$ of the transverse
carrier motion (the latter is about 16 meV at $n=n_*$ and increases to 
$E_0-E_{cs} \approx 47$ meV at $n=10^{12}$ cm$^{-2}$, whereas $nU_{2D}$ increases
from $0.05$ meV to about 1 meV). In this regime, the short-range $U$ 
almost does not perturb the transverse motion, and in particular the 
magnetisation dependence of the carrier wave functions 
(see Fig.  \ref{fig:scheme}) is very weak. In turn, this magnetisation 
dependence of $\psi_\alpha(z)$ is the only ingredient that distinguishes
the full numerical solution of mean-field equations from the ``Stoner'' approach
which yields Eq.(\ref{eq:2dstonertilde}).

The dotted line in Fig. \ref{fig:dolgo} corresponds to using the  {\it ansatz}, 
Eq. (\ref{eq:Stern}), for $\psi_\alpha(z)$, which amounts to substituting 
$U_{2D}^{var}$ for $U_{2D}$ in Eq. (\ref{eq:2dstonertilde}). This would
slightly underestimate the  value of density at the ferromagnetic transition, 
the discrepancy being due
to the variational nature of this approach.

The ferromagnetic transition is second-order, and the full polarisation is
reached at a certain density $n_F<n_*$. Numerically, we find that the transition
is very steep, with
$n_*-n_F \sim 5 \cdot 10^7$ cm$^{-2}$. The latter value presumably is well 
below any 
experimental accuracy. This is in line with the preceding discussion: 
as explained in Sect. \ref{sec:single} above, within the conventional
Stoner approach the mean-field
transition would have been perfectly abrupt. The fact that the transition is
in fact smooth is due to the dependence of wave functions on $M$ 
(Fig.  \ref{fig:scheme}), which is very weak at low densities. Indeed, in the 
fully
polarised state below $n_F$ we find\footnote{Here $n_\downarrow=0$, and we need 
to re-define $z_\downarrow$ as 
$z_\downarrow=\int \psi_{0,\downarrow}^2(z)z dz        $ 
[cf Eq.(\ref{eq:zalpha})].}
%\cite{zdown} 
$(z_\downarrow-z_\uparrow)/(z_\downarrow+z_\uparrow) \approx 6 \cdot 10^{-4}$, reflecting a rather minute difference in the
profile of spin-up and spin-down distributions. This should be contrasted with
a pronounced difference between $z_\uparrow$ and $z_\downarrow$ above the 
high-density magnetic transition, as seen in Fig. \ref{fig:U2D}.

We emphasise that this mean field picture may be significantly modified once
the effects of fluctuations are taken into account\cite{Conduit2013}. These
may increase the value of $n_*-n_c$ and turn the transition first-order;
the latter would be in line with reported inhomogeneous behaviour in this
region\cite{Teneh12}.

The effects of finite temperature (beyond the strictly degenerate regime) are 
outside the scope of the present article. We speculate that the peak (rather 
than a divergence) of the effective mass reported in Ref. \cite{Kuntsevich15} 
may correspond to the scenario whereby the ferromagnetic ordering is stabilised
at temperatures below those used in Ref. \cite{Kuntsevich15}.  

Our tentative results as outlined above imply that a ferromagnetic transition occurs
at a critical value of density $n_*$ which is a few per cent larger than that
of the MIT ($n_c$). On the other hand, available experimental results suggest 
the following two scenarios: (i) As the density is decreased toward the MIT,
the susceptibility increases, reaching a large but finite value at the point
of MIT\cite{Reznikov03}. Then the (asymptotic) value of transition critical 
density $n_*$ would lie below $n_c$ (the ferromagnetic transition
is preempted by the MIT, at which point the properties of the system change and
there is no transition at $n=n_*$). (ii) The susceptibility actually diverges 
in the vicinity of the MIT, with the
two transitions occurring simultaneously or very close to 
each other\cite{Kravchenko04}\footnote{In addition to susceptibility 
measurements, further
support comes from the density dependence of magnetic field value required
to fully spin-polarise the system\cite{Shashkin01,Vitkalov01}.}.
%,Shashkin01}. 
While 
it might appear that our present 
conclusions do not support either of these two possibilities, we wish to argue 
that our results can be re-interpreted and reconciled with the second one.

Once the system is fully spin-polarised by an applied field, it 
exhibits insulating behaviour even at densities {\it above} the $H=0$ MIT 
point\cite{Kravchenko04}. The
in-plane field can affect transport properties only via spin, {i.\ e.} via the 
magnetisation $M$ (or equivalently via the degree of spin polarisation). Thus,
it seems logical to expect that whenever the system is fully spin-polarised 
(either due to an external field or to intrinsic ferromagnetism), it turns
insulating. That would mean that the actual MIT takes place at $n=n_*$ 
(we recall that the width of magnetic transition is expected to be 
negligible), whereas $n_c$ (which is a few percentage points below $n_*$) 
retains the 
meaning of
an extrapolation parameter controlling the bandwidth renormalisation 
[see Eqs.(\ref{eq:dolgo}--\ref{eq:nutilde})]. 
%It further appears that the 
%accuracy of experimental results on effective mass renormalisation, cited in
%Ref. \onlinecite{Dolgopolov15}, allows for such a re-interpretation.  
We note 
that the latter is somewhat similar to the scenario discussed in Ref. 
\cite{ZhangDasSarma2005} in the context of long-range Coulomb 
interaction alone.

The available experimental data for the effective mass (which can be
deduced, {\it e.g.}, from the transport measurements\cite{Shashkin02}) and
susceptibility do not allow to conclude with certainty that the latter indeed 
follows 
either Eq. (\ref{eq:2dstonertilde}), and not
Eq. (\ref{eq:Pauli}). The observed 
systematic 
differences\cite{Kravchenko06,Kravchenko06-2,Dolgopolov15} 
[see, {\it e.g.}, Fig. 9 in Ref.
\cite{Kravchenko06}] between the 
measured $\chi$ and the calculated Pauli value $\chi_P$ [Eq. (\ref{eq:Pauli})] 
may be due, at least in part, to the experimental issues or inaccuracies of
interpretation. In order to reliably verify the importance of short-range 
interaction, further measurements would need to be performed closer to
the MIT.

\section{The high-density metallic regime}
\label{sec:metal}

\begin{figure}  \sidecaption
\includegraphics[width=.49\textwidth]{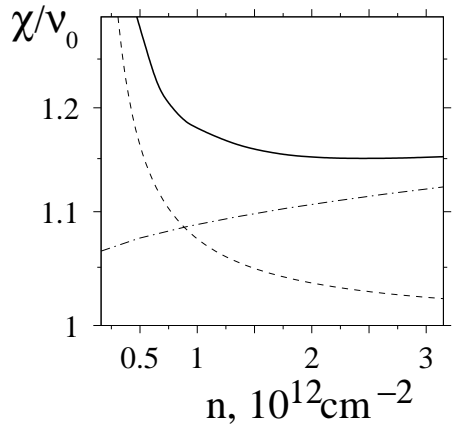}
\caption{\label{fig:cross} Magnetic susceptibility $\chi$ divided by
  the bare density of states $\nu_0$ in the intermediate density range.
  Solid and dashed  lines show, respectively,  
  the numerical solution of the mean-field  equations [taking into account
the effective mass renormalisation,  Eq.(\ref{eq:dolgo})] and the 
Pauli susceptibility, Eq.  (\ref{eq:Pauli}). 
Dashed-dotted line corresponds to the numerical solution of the mean-field 
equations with the unrenormalised density of states $\nu_0$.} 
\end{figure}

When the density is increasing further away from the MIT, the susceptibility
continues to decrease, as shown by the solid line in Fig. \ref{fig:cross}
(which is a continuation of the solid line in Fig. \ref{fig:dolgo}). This is
due to the decreasing influence of the long-range correlations 
[taken into account phenomenologically via Eq. (\ref{eq:dolgo})], and indeed
reflects the decreasing $U=0$ phenomenological susceptibility [Pauli 
susceptibility, Eq. (\ref{eq:Pauli}), dashed line in Figs. \ref{fig:dolgo} and
\ref{fig:cross}]. Qualitative estimate confirms that in this region the 
long-range correlations weaken and ultimately cease to dominate, with 
$r_s \approx 1$ at $n=3\cdot 10^{12}$ cm$^{-2}$. It is seen that as the value
of $n$ continues to increase, the susceptibility passes through a broad minimum
at $n \approx 2.4\cdot 10^{12}$ cm$^{-2}$ and begins to increase. The latter 
feature is due to the
increasing role of the contact interaction $U$. This corresponds to
the increase shown by the dashed-dotted line, which depicts the value of 
susceptibility calculated using the unrenormalised value $\nu_0$ 
[see Eq. (\ref{eq:nu0})] of the density
of states (i.\ e.,  shows the effects of $U$ only, along the lines of 
Sect. \ref{sec:single}).

A further increase in $n$ leads to populating the second (first excited)
level of carrier motion in the $z$ direction. Indeed, we already mentioned
in Sect. \ref{sec:single} that modelling the behaviour of the system at
higher inversion-level carrier densities $n$ requires taking into account the 
presence of multiple occupied levels. Thus, one has
to implement the complete mean field scheme, without a simplification utilised 
in Sects. \ref{sec:single} and \ref{sec:dolgo}, where we used Eq. (\ref{eq:nalpha0}) in place of a more general
Eq. (\ref{eq:nalpha}). On the other hand, at these larger values of $n$ the 
phenomenological carrier density of states $\tilde{\nu}$ 
[see Eq.(\ref{eq:nutilde})]
approaches its unrenormalised value $\nu_0$. Indeed, the effect of the mass
renormalisation at $n=3\cdot 10^{12}$ cm$^{-2}$ on susceptibility is already 
negligible (the difference between solid and dashed-dotted lines at the right
edge of Fig. 
\ref{fig:cross}), and decreases further with increasing $n$. Thus, we cross 
into the normal Fermi liquid regime, and we may use  the unrenormalised 
value $\nu_0$ of the density of states (which somewhat simplifies the 
complicated numerical calculation). A possibility of strong 
{\it Fermi-liquid} renormalisations at larger $n$ owing to the  
contact interaction $U$ will be discussed in Sect. \ref{sec:conclu}. 
Given the absence
of data for high densities, we will be using the unrenormalised 
value $\nu_0$ throughout.

We again begin with the conventional Stoner mean-field description of the 
paramagnetic phase, assuming that the transverse wave functions 
$\psi_{l,a,\alpha}(z)$ do not change when the magnetisation $M$ varies. The 
latter
assumption is essentially a variational one, and implies that when $M$ is 
small, 
the $(l,a)$th transverse energy level of a spin-up 
electron (which at $M=0$ is given by $E_{l,a,\uparrow}$) acquires a correction,
\begin{equation}
\delta^{(0)} E_{l,a,\uparrow}=\sum_{l'\!,a'\!} U_{2D}^{l,a;l'\!,a'\!} \delta^{(0)}
n_{l'\!,a'\!,\downarrow}-\frac{1}{2}H \,,
\label{eq:multishifts}
\end{equation}
and similarly for spin-down electrons. The matrix $U_{2D}$ 
(which in the paramagnetic phase is symmetric) is defined by Eq. (\ref{eq:U2D}),
and the corrections $\delta^{(0)}n_{l,a,\alpha}$ to the level occupancies at $M \neq 0$ are found self-consistently from Eq. (\ref{eq:nalpha}). This leads to a 
set of
self-consistency equations,
\begin{eqnarray}
&&\delta^{(0)} n_{l,a,\uparrow}-\delta^{(0)} n_{l,a,\downarrow}= 
\gamma_l \nu_l \times \nonumber \\
&& \times\left[H+ \sum_{l'\!,a'\!} 
 U_{2D}^{l,a,l'\!,a'\!} \left(\delta^{(0)} n_{l'\!,a'\!,\uparrow}-\delta^{(0)} n_{l'\!,a'\!,\downarrow} 
\right)\right].
\label{eq:stonersystem}
\end{eqnarray}
This linear system is readily solved, and the ``Stoner'' susceptibility is then 
found as
\begin{equation}
\chi_0=\frac{1}{2H}\sum_a\left(\delta^{(0)} n_{a,\uparrow}-\delta^{(0)} 
n_{a,\downarrow}\right)\,.
\label{eq:2dstonermulti}
\end{equation}
In the single-level case, 
Eq. (\ref{eq:2dstonermulti}) yields the familiar single-level result,
 Eq. (\ref{eq:2dstoner}). On the other hand, we note that in the multi-level 
case the
``Stoner'' result (\ref{eq:2dstonermulti}) includes effects of the restricted
geometry, not found in either 3D bulk or purely 2D systems (see below).

\begin{figure} \sidecaption
\includegraphics[width=.49\textwidth]{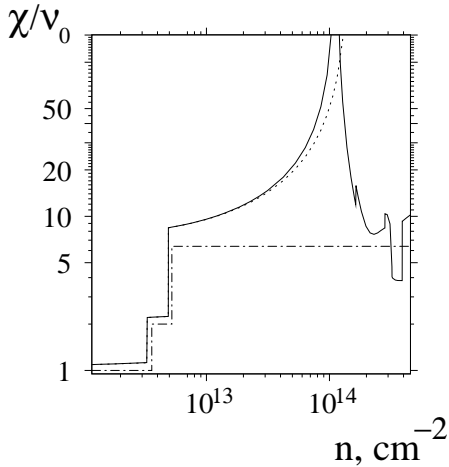}
\caption{\label{fig:chimulti} Magnetic susceptibility $\chi$ 
at large inversion-layer carrier densities $n$. Solid line depicts the 
numerical results of 
the full mean-field multi-level calculation, showing
transition at  $n_{FM}\approx 1.15 \cdot 10^{14}$
cm$^{-2}$. Dotted line corresponds to the Stoner value $\chi_0$, as 
derived from Eq. 
(\ref{eq:2dstonermulti}), and the dashed-dotted line represents the
non-interacting case of $U=0$. } 
\end{figure}

As long as the carrier density is not too high, $n \stackrel{<}{\sim} 3 \cdot 10^{13}$ cm$^{-2}$, 
the susceptibility value 
obtained by numerically solving the mean field equations (solid line in Fig.
\ref{fig:chimulti}) is well described by the Stoner theory [dotted line,
obtained from Eq. (\ref{eq:2dstonermulti})]. As expected already in the
non-interacting case ($U=0$, corresponding to the dashed-dotted line in Fig.
\ref{fig:chimulti}), once a new transverse motion level is populated 
the
susceptibility suffers a jump. For our values of parameters we find that
these are located at
$n^{(0,1)}\approx 3.3 \cdot 10^{12}$ cm$^{-2}$ and 
$n^{(1,0)}\approx 4.9 \cdot 10^{12}$ cm$^{-2}$ (where the superscript is the 
number
of the transverse motion level which dips below the Fermi level at the 
corresponding value of $n$, preceded by the number of the corresponding 
ladder). 
We note that the magnitude of
the steps in $\chi/\nu_0$ is renormalised in comparison with the non-interacting
case, where for a step at every $n=n^{(l,a)}$ we find 
$\delta \chi^{(l,a)}=\gamma_l\nu_l/2$.
For example, the magnitude of the step at $n=n^{(0,1)}$ in our case is 
$\delta \chi/\nu_0\approx 1.08$. As readily seen with the help of  
Eq. (\ref{eq:2dstonermulti}), the difference from unity is due to the non-zero
matrix
elements $U_{2D}^{0,0;0,1}=U_{2D}^{0,1;0,0} (\approx 2.7 \cdot 10^{-28}  
{\rm erg \cdot cm}^2)$ and $U_{2D}^{0,1;0,1} 
(\approx  9.9 \cdot 10^{-28}{\rm erg \cdot cm}^2)$. On the other hand, 
the difference of the baseline
value of $\chi/\nu_0$ just below the step,  $\chi/\nu_0\approx 1.12$, 
from unity is due to the (larger) $U_{2D}^{0,0;0,0} (\approx  2.2 \cdot 10^{-27} 
{\rm erg \cdot cm}^2)$. Note that
in this density range, the values of $U_{2D}^{l,a;l\,',a\,'}$ at fixed $n$ are 
approximately proportional to $U$.  

At $n \sim n^{(0,1)}$, the magnetic susceptibility $\chi$ deviates only 
slightly 
from
its value in the non-interacting case (see the dashed-dotted line in Fig.
\ref{fig:chimulti}), confirming that the effects of 
short-range interaction are relatively weak. Thus it is natural that the
precise value of $ n^{(1,0)}$ does not strongly depend on $U$, {\it e.g.,} at
$U=0$ we get\cite{Stern72} $ n^{(0,1)}=3.6\cdot 10^{12}$ cm$^{-2}$. On the other hand,
$ n^{(0,1)}$ is sensitive to the acceptor density $N_A$ which can vary broadly.
Indeed, for $U=0$ and $N_A=10^{14}$ cm$^{-3}$ we find 
$n^{(0,1)}\approx 2.2\cdot 10^{12}$ cm$^{-2}$ (which again agrees with Ref. 
\cite{Stern72}), whereas for $U=0$ and $N_A=10^{16}$ cm$^{-3}$,
$n^{(0,1)} \approx 6.1\cdot 10^{12}$ cm$^{-2}$. 

With a further increase in density, the numerical results for $\chi$ in Fig.
\ref{fig:chimulti} begin to
deviate from the Stoner susceptibility $\chi_0$. This is because the 
short-range interaction
begins to affect the transverse carrier motion, and the wave functions become
polarisation dependent (see Fig. \ref{fig:scheme}). Indeed, at 
$n =3 \cdot 10^{13}$ cm$^{-2}$ the most important energy scale of 
the
transverse motion, $E_{0,0,\alpha}-E_{cs} \approx 480$ meV, is only a few times 
larger than
the (roughly estimated) interaction energy scale, 
$2n_{0,0,\uparrow}U_{2D}^{0,0;0,0} \sim 
70$ meV 
(see discussion in Sect. \ref{sec:dolgo} above).

This deviation of $\chi$ from $\chi_0$ further increases 
with $n$, until $\chi(n)$ becomes critical signalling a second-order 
ferromagnetic phase transition at $n_{FM}\approx 1.15 \cdot 10^{14}$ cm$^{-2}$.
As mentioned above, at this point carriers populate three spin-degenerate 
levels of the
$z$-axis motion, which is the reason behind the increase in the critical
density  $n_{FM}$ in comparison to the single-level estimate $n_1$ of Sect. 
\ref{sec:single}. Indeed, the wavefunction of higher levels are broader in the
$z$-direction, which results in smaller values of the corresponding 
$U^{i,j}_{2D}$ (see the data for $n=n^{(0,1)}$ above) and hence in a certain 
weakening of the interaction effects.

Overall, the dotted line in Fig. \ref{fig:chimulti}, which shows the 
multi-band Stoner 
susceptibility $\chi_0$, Eq. (\ref{eq:2dstonermulti}), 
%for the susceptibility in 
%the paramagnetic state (based on the numerically calculated values of 
%$U_{2D}^{l,a;l',a'}$). We note that here $\chi_0$ 
follows the numerical 
result much more closely than in the single-level case of 
Sect. \ref{sec:single}
(see Fig. \ref{fig:susce}). The reason is that, as mentioned above, 
the respective
transverse wavefunction spreads differ  for different active levels. 
Within
the multi-level Stoner scheme, at $H \neq 0$ these levels 
are shifted in a {\it non-uniform} self-consistent fashion 
[see Eq. \ref{eq:multishifts})], giving rise to an 
$H$-dependent difference in the profile of the net spin-down and spin-up charge 
densities (cf. Fig. \ref{fig:scheme}). In this way, a Stoner treatment 
yielding Eq.  
(\ref{eq:2dstonermulti}) is able, in the multi-level case only, to partially
mimic the effect of wavefunction change as captured by the full numerical 
solution of the mean field equations, resulting in a better fit. 

Still, we find that Eq. (\ref{eq:2dstonermulti}) predicts\footnote{Values of 
$U_{2D}^{l,a;l',a'}$, needed to evaluate $\chi_0$ 
in the region 
$n_{FM} < n < n_0$, are 
obtained by finding the $M=0$ (spin-degenerate) solution to the mean field 
equations, even as this solution does not minimise the thermodynamic
potential, Eq. (\ref{eq:G}).}
%\cite{calculU} 
a (discontinuous) ferromagnetic 
transition at $n_0 \approx 1.47 \cdot 10^{14}$ cm$^{-2}$, well above the 
actual transition density $n_{FM}$.  
Hence the adequate self-consistent treatment of 
the wave function dependence on $M$ is important for  
evaluating  the critical density. In a direct analogy with Sect. 
\ref{sec:single}, we conclude
that the Stoner criterion of ferromagnetism 
is {\it relaxed}.

\begin{figure}  \sidecaption
\includegraphics[width=.49\textwidth]{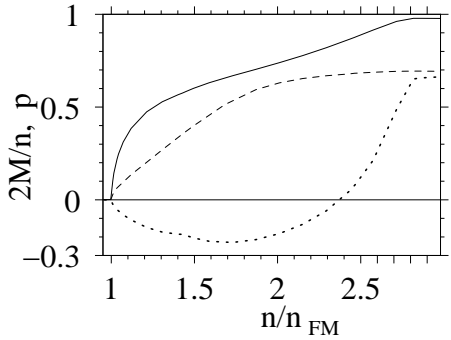}
\caption{\label{fig:Mmulti}  Numerical results for the 
degree of spin polarisation, $2M/n$, at the densities $n$ above the 
ferromagnetic transition: $n>n_{FM} \approx  1.15 \cdot 10^{14}$ cm$^{-2}$ 
(solid line).
The dashed and dotted lines shows the values of $p^{(1,0)}$ and $p$, Eqs. 
(\ref{eq:p10}) and (\ref{eq:p}), respectively. These highlight the
difference between spin-up and spin-down carrier distribution along the 
$z$-axis.}
\end{figure}

Density dependence  of the spontaneous magnetisation, $M(n)$, is shown in
Fig. \ref{fig:Mmulti} (solid line). The fact that the transition at $n=n_{FM}$ 
is smooth
is explained (as in Sect. \ref{sec:single}, see also Sect. \ref{sec:field}) 
by the magnetisation dependence
of the $z$-axis motion wavefunctions. This effect is surprisingly strong: an
increase of $n$ by a factor of 2.8 is required to saturate the relative 
magnetisation.
Interestingly, the value of $2M/n$ then reaches a plateau at about 0.98 (with 
the 0th 
spin-down levels in both ladders pinned just below the Fermi energy). The 
complete spin polarisation, $M=n/2$, is not attained even at  $n \sim 2.3 \cdot 
10^{15}$
cm$^{-2}$. Given the inversion layer thickness of the order of $10^{-7}$ cm,
this value approaches the normal-metal range of three-dimensional carrier 
densities, where our approach becomes invalid.

Owing to a larger effective mass and higher valley degeneracy, the only 
active level (0th) in the 1st ladder  provides most of the density of states at the 
Fermi level. The evolution of average $z$ values of carriers in this level,
\[z^{(1,0)}_\alpha = \int \psi_{1,0,\alpha}^2(z)z dz
\,,\]
with $M$, is characterised by increasing ratio
\begin{equation}
p^{(1,0)}=\frac{z^{(1,0)}_\downarrow-z^{(1,0)}_\uparrow}{z^{(1,0)}_\downarrow+z^{(1,0)}_\uparrow}
\label{eq:p10}
\end{equation}
(dashed line in Fig.\ref{fig:Mmulti}). Clearly, the spatial separation 
between opposite-spin carriers belonging to this level
increases with magnetisation, and the magnitude of $p^{(1,0)}$ mirrors 
the value of $M$.
This can be understood in terms of Fig. \ref{fig:scheme} (see discussion in 
Sect. \ref{sec:single}). 

On the other hand, the 
behaviour of overall average values $z_\alpha$ [including 
contributions from all active levels, see Eq. (\ref{eq:zalpha})] is 
complicated by effects 
of particle
re-distribution between different levels, as well as by inter-level interaction. For example, as $M$ increases, a
larger fraction of minority carriers resides in the levels of the 0th ladder  
which may {\it reduce} the ratio
\begin{equation}
p=\frac{z_\downarrow-z_\uparrow}{z_\downarrow+z_\uparrow}\,
\label{eq:p}
\end{equation}
(see the dotted line in Fig. \ref{fig:Mmulti}).

%The fact that the $z$-axis wave functions for spin-up and spin-down
%carriers (which are used also for calculating $U_{2D}^{i,j}$) {\it are} 
%different for $n>n_{FM}$ implies that Eq.
%(\ref{eq:2dstonermulti}) cannot be used in the $M \neq 0$ case.
%Hence the dashed line in Fig. \ref{fig:chimulti}, showing the value of $\chi_0(%n)$ terminates at $n=n_{FM}$.
%Note that while increasing with $n$, the Stoner theory result $\chi_0$ shows
%no sign of critical behaviour.
%Hence the adequate self-consistent treatment of 
%the wave function dependence on $M$ is qualitatively important for  
%evaluating  $n_{FM}$. We refer to Sec. \ref{sec:single} for the discussion of 
%single-level calculation, where the comparison with variational results is
%presented. 

%In a direct analogy with Sec. \ref{sec:single}, we conclude
%that the Stoner criterion of ferromagnetism [which in
%the multi-level case corresponds to the vanishing determinant ${\cal D}$ 
%of the equations system (\ref{eq:stonersystem})] 
%is {\it relaxed}. Indeed, for a given
%density the value of $U$ required to destabilise the paramagnetic state
%is {\it smaller} than the one implied by the ${\cal D}=0$ condition.

In the ferromagnetic phase, spin-up and spin-down carriers no longer 
begin to populate new $z$-axis motion levels simultaneously. Indeed, our
result for $\chi(n)$ shows further upward steps at 
$n^{(0,2)}_\uparrow\approx 1.64 \cdot 10^{14}$ cm$^{-2}$,
$n^{(0,3)}_\uparrow\approx 2.83 \cdot 10^{14}$ cm$^{-2}$, and
$n^{(1,1)}_\uparrow\approx 3.91 \cdot 10^{14}$ cm$^{-2}$,
 where spin-up electrons 
(only) begin to populate the 2nd and 3rd excited levels in the 0th ladder and 
the 1st excited level in the 1st ladder, respectively. In addition, there is
a downward step at $\tilde{n}^{(0,1)}_\downarrow\approx 3.23 \cdot 10^{14}$ 
cm$^{-2}$, where due to increasing polarisation $M(n)$, the spin-down electrons
cease to populate the 1st excited level in the 0th ladder. Interestingly, these
points do not correspond to any noticeable features of magnetisation, $M(n)$ 
(see Fig. \ref{fig:Mmulti}). 
Overall, the non-monotonous
density dependence of  $\chi(n)$ in Fig. \ref{fig:chimulti} in the 
ferromagnetic region above $n_{FM}$  should be ascribed to a 
combined effect of the wave functions changing and the carriers 
redistributing between the 
bands with increasing $M$.   

While relegating further discussion of these results to Sect. \ref{sec:conclu},
we note that solving the mean-field equations in the multilevel case, in a broad
range of values of density $n$, is a delicate numerical problem. For a given 
value of $E_{cs}$, the mean field
equations  (see Sect. \ref{sec:mf}) are first solved for a suitable variational 
ansatz of the type (\ref{eq:Stern}), yielding the values of $z_d$, $n$, and $M$ and the 
corrected wave functions; these are then fed back into the mean field equations
and the process repeated until convergence is achieved 
(cf. Ref. \cite{Stern70}).  It is found that the value of $M$ converges
rather slowly (as opposed to $n$ and $z_d$), necessitating a large number of
iterations (up to some 8400 near the critical point, $n=n_{FM}$). 
In addition, since 
the wave 
function spread in the $z$-direction increases for higher levels, particular 
care should be taken in choosing large-$z$ cutoff $z_{max}$ when solving the 
Schr\"{o}dinger equation (\ref{eq:Schroed}) and evaluating required integrals. 
For the values of $n$ shown in Fig. \ref{fig:chimulti}, we found it necessary
to increase the ratio of $z_{max}$ to the average carrier coordinate $(z_\uparrow n_\uparrow + z_\downarrow n_\downarrow)/n$ in stages from 7 for smaller $n$ to 
34 for largest values. This subtlety, as well as the important role played by
the $l=1$ ladder of energy levels, was overlooked in Ref. 
\cite{jmmm}, hence the preliminary results for the high-density regime reported
therein are quantitatively incorrect.
%\cite{correct}.

\section{Sublinear magnetisation}
\label{sec:field}

In a purely 2D system, Stoner approach yields the value of  magnetisation 
$M(H)$ which increases linearly with field from
$H=0$ all the way up to the saturation field $H_s$. This is a consequence of
the 2D density of states being energy-independent. When several 2D bands are
present (corresponding in our case to different ladder and level indices 
$l,a$), the complete spin polarisation within a given band may be attained 
at field values $H_{l,a}^\downarrow<H_s$, corresponding to $E_{l,\alpha,\downarrow}>0$ [cf. Eq. (\ref{eq:nalpha})]. In addition, new bands $l',a'$ may become 
available as the corresponding energy for spin-up particles drops below the 
chemical potential ($E_{l',\alpha',\uparrow}<0$); we denote the corresponding 
fields $H^\uparrow_{l',a'}$.
The value of dynamic susceptibility $\chi(H) \equiv \partial M /\partial H$
then shows jumps at $H^{\uparrow,\downarrow}_{l,a}$, while remaining
constant elsewhere. These constant values of $\chi(H)$ between the jumps depend
on the thermodynamic formulation of the problem -- whether it corresponds
to the chemical potential (more precisely, $\mu-E_{cs}$) or net carrier 
density $n$ being fixed. As mentioned
in Sect. \ref{sec:mf}, our system is closer to the latter regime (see below).

\begin{figure} \sidecaption
\includegraphics[width=.49\textwidth]{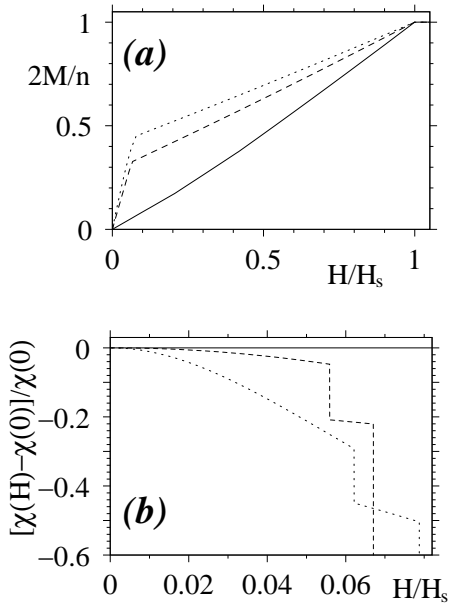}
\caption{\label{fig:MH} (a) Numerical results for the 
degree of spin polarisation in the paramagnetic phase, $2M/n$, plotted
as a function of renormalised magnetic field $H/H_S(n)$, where $H_s(n)$ is the
saturation field. Solid, dashed, and dotted line correspond, respectively, to
the following $H=0$ values of density $n$: $2.63 \cdot 10^{12}$ cm$^{-2}$  
($H_s \approx 25$ meV in energy units), $3.14 \cdot 10^{13}$ cm$^{-2}$ 
($H_s \approx 139 $ meV), and 
$6.62 \cdot 10^{13}$ cm$^{-2}$ ($H_s \approx 198$ meV). (b) The relative
change of magnetic susceptibility $\chi$ at low fields for the same
values of $n(H=0)$.}
\end{figure}

The results of numerical calculation of $M(H)$ for our system at three 
different $H=0$ carrier densities in the metallic regime are shown in Fig. 
\ref{fig:MH} (a). We see that $M(H)$ increases monotonically and continuously
all the way up to saturation; there is no evidence of a discontinuity at 
$H=H_s$, which was reported\cite{DasSarma2006,Subasi2008} 
in the case of a 2DEG with Coulomb repulsion. For higher densities, one
observes pairs of features (cusps), merged together on the scale of the figure.
For $n=3.14 \cdot 10^{13}$ cm$^{-2}$, these correspond to 
$H^\downarrow_{0,1}/H_s \approx 0.056$ and  
$H^\downarrow_{1,0}/H_s \approx 0.067 $, 
whereas for  
$n= 6.62 \cdot 10^{13}$ cm$^{-2}$ we find  $H^\downarrow_{0,1}/H_s \approx 0.062 $
and $H^\downarrow_{1,0}/H_s \approx 0.079$ (owing to a larger combined density 
of states $\gamma_1 \nu_1$, $H^\downarrow_{1,0}$ corresponds to a stronger feature). 
For   $n =2.63 \cdot 10^{12}$ cm$^{-2}$, there are two weak barely visible 
features 
corresponding to  $H^\uparrow_{0,1} \approx 0.20  H_s$ and  
$H^\uparrow_{1,0} \approx 0.42  H_s$.

As explained in Sect. \ref{sec:mf} our calculation is performed at a fixed 
value of the gate voltage $\phi_{gate}$, thus modelling the actual experimental 
setup. We find that for the zero-field density $n= 6.62 \cdot 10^{13}$ 
cm$^{-2}$ increasing value of $H$ from 0 to $H_s$ leads to a 
decrease of the absolute value of $E_{cs}$ by some 5\%, whereas the density $n$
{\it increases} by about $0.003$ \%. Corresponding values for the other
two curves on Fig. {\ref{fig:MH}} are similar. We see that indeed the 
system is much
closer to the fixed-$n$ regime than to that of a constant $E_{cs}$.
 We note that in all cases, the value of magnetic 
length\footnote{Note that our $H$ is defined in the units of Bohr magnetone.}
%\cite{lb}
\begin{equation}
l_B=\left(\frac{\hbar c \mu_B}{eH}\right)^{1/2}=\hbar/\sqrt{2m_eH}
\end{equation}
at $H=H_s$ is two to three times smaller than the average value of $z$ for the
carriers, suggesting the importance of orbital effects of the in-plane field. 
While we do not take these effects into account, we note that 
elsewhere\cite{Tutuc03} these were found to result in a slight upward bend 
(superlinear behaviour) of 
the $M(H)$ curve at $U=0$ at low densities.

It may appear that the behaviour of $M(H)$ as shown in Fig. \ref{fig:MH} (a)
is linear except for the features at $H_{l,a}^{\uparrow,\downarrow}$. In reality,
this holds only for the lowest density,  $n =2.63 \cdot 10^{12}$ cm$^{-2}$, 
where the effects of short-range interaction are too weak to affect the
transverse carrier motion. This is illustrated  by
Fig.  \ref{fig:MH} (b), depicting relative change of the dynamic 
susceptibility with $H$ at low fields\footnote{At these low fields, 
our omitting the orbital effects is 
mathematically justified.}
%\cite{lowfield}
. The pairs of 
susceptibility jumps at $H=H^\downarrow_{0,1}$ and $H=H^\downarrow_{1,0}$ are seen
for higher densities. In addition, the appreciable decrease of $\chi(H)$ with
$H$ at $H<H^\downarrow_{0,1}$ implies a {\it sublinear} magnetic field 
dependence of $M$ in this region.
This behaviour becomes more pronounced as the density increases toward the
ferromagnetic instability. 

This sublinear behaviour of magnetisation is due to the effect of carrier wave 
functions changing with increasing $M$, as discussed above. Indeed, the 
effective interaction $U_{2D}$ enhances
the magnetic susceptibility in comparison to its non-interacting value.
 With increasing $M$, the wave-function profiles are adjusted in such a way 
that the interaction energy is lowered. Hence the effective value of $U_{2D}$ 
decreases (see Sect. \ref{sec:mf}) and so does the susceptibility.

The sublinear field dependence of $M$ is of  crucial importance for one 
feature of the
present theory which is not expected in the conventional Stoner treatment of a 
purely 2D system, {\it viz.}, the continuous character of ferromagnetic 
transitions (Sects. \ref{sec:single}, \ref{sec:dolgo}, \ref{sec:metal}).
Indeed, a simple Landau--Ginzburg type description implies that a continuous
transition requires the presence of a {\it positive} quartic (in $H$) term
in the free energy, and hence sublinear magnetisation.
It is hoped that perfecting the magnetisation measurement techniques and 
extending them to the
higher-density region (where the long-range correlations become negligible) 
will allow to directly confirm this behaviour in a Si-MOSFET.

\section{Conclusion}
\label{sec:conclu}

We constructed a mean-field description of  electrons in an 
inversion layer, addressing both the behaviour of the system in the metallic 
high-density region and
the correlated low-density regime immediately above the metal-insulator 
transition. Such electronic systems [as exemplified by Si-(100) MOSFETs] 
are characterised by the presence of both long-range Coulomb repulsion
and the ubiquitous short-range (on-site, Hubbard) interaction. Coulomb
interaction was treated at the mean-field level following Ref. 
\cite{Stern72},
which leaves out long-range correlation effects, important in the low-density
limit where the dimensionless parameter $r_s=m_\bot e^2/(\epsilon \hbar^2 
\sqrt{\pi n})$ is large (we included these
effects phenomenologically in Sect. \ref{sec:dolgo}). 

We recall that in a bulk three-dimensional system of electrons interacting 
via contact 
potential, Eq. (\ref{eq:contact3D}), the strength of this interaction is
measured by the dimensionless quantity $k_F a_{3D}$, where $k_F$ is the Fermi
wave vector and $a_{3D} = m_*U/4 \pi \hbar^2$  (where $m_*$ is the appropriate 
3D effective mass\footnote{Here, the scattering length is defined using the
{\it reduced} mass of a pair of identical particles, {\it i.e.}, 
in this case, $m_*/2$.}
%\cite{reduced}
) is the scatterring length in the
Born approximation.
Therefore one expects that in a dilute system (small $k_Fa_{3D}$, large $r_s$) 
the effects of short-range interaction are negligible. At larger densities, 
the increasing value of $k_F a_{3D}$ gives rise to stronger Fermi-liquid 
renormalisations (in particular, enhancing the magnetic susceptibility); at 
the same time, smaller values of $r_s$ ($\propto m_* e^2/
\epsilon \hbar^2 n^{1/3}$ in the three-dimensional case) and enhanced 
screening eventually permit neglecting the 
long-range Coulombic correlations. Depending on the properties of the system,
it may or may not  undergo a Stoner transition, accompanied by a 
susceptibility divergence. 

In a restricted geometry of an inversion layer (a quasi-2D system), this 
picture is modified in a drastic way. Momentum dependence of the $s$-wave 
scatterring amplitude in 2D 
(see, {\it e.g.,} Ref. \cite{Landau3}) 
yields the momentum-dependent scatterring length\cite{Verhaar84} $a_{2D}(k)$. 
Assuming for simplicity that only one level of transverse motion is active
(electrical quantum limit, Sect. \ref{sec:single} above), one finds\footnote{
Eq. (\ref{eq:a2d}) is 
obtained 
from the requirement\cite{Verhaar84} that  scattering phase shifts 
for the contact and 
hard-core potentials coincide.}
%\cite{phaseshift} 
for a 
given 2D wave vector $k$, in the Born approximation\footnote{Which is expected to provide a valid estimate throughout our
range of values of parameters, see Appendix.}
%\cite{Born} 
for the contact 
interaction, 
Eq.(\ref{eq:U2D}): 
\begin{equation}
\log \frac{2}{ka_{2D}} =\frac{2 \pi \hbar^2}{m_\bot U_{2D}}+ \gamma_E\,. 
\label{eq:a2d}
\end{equation}
Here, $\gamma_E \approx 0.577 $ is Euler's constant.  The (short-range) 
interaction strength parameter in the 2D case is given 
by\cite{Randeria,Randeria2} 
$g=[\log(2/k_F a_{2D}(k_F))]^{-1}$. According to Eq. (\ref{eq:a2d}), 
in the absence
of long-range correlations the value of $g$ depends on the 2D 
density $n$ only via $U_{2D}$. While the latter does grow with $n$ owing to
decreasing inversion layer thickness (the latter, as dictated by electrostatics), this
growth is relatively slow (see Fig. \ref{fig:U2D}). Indeed, we estimate that 
as the density varies from $8 \cdot 10^{10}$ cm$^{-2}$ to $8 \cdot 10^{13}$ 
cm$^{-2}$, the value of $g$ increases from about  $0.03$  to $0.12$. 
This increase, implying an appreciable effect of interaction at larger 
densities, is expected to be  more pronounced in a real multi-level system,
where the valleys with larger in-plane mass are populated.

In a 2D system where there is no coexistence of broad and narrow 
partially-filled bands at the Fermi level, the short-range interaction is 
generally not expected to easily yield ferromagnetism (as exemplified by the 
square-lattice
Hubbard model, see, {\it e.g.}, Refs. \cite{Dombre,Edwards,Wurth}). 
Even in the range of densities 
where the
ferromagnetism does occur, the required interaction strength is so large that
Stoner mean-field approach is clearly irrelevant 
(see, however, Ref. \cite{Conduit2013}). However, in the case of a 
silicon inversion layer at high densities, there is an additional mechanism
(transverse wave function dependence on magnetisation) acting alongside the
conventional Stoner one ({\it viz.,} the mean-field shifts of band
energies). This opens an additional avenue toward ferromagnetic instability in 
the range
where mean field approach is still expected to be applicable (see Appendix).
In the opposite case of very low densities just above the MIT, the 
interaction-induced wave-function changes are negligible, yet there is a strong
renormalisation of carrier properties due to the long-range Coulomb 
correlations\cite{Finkelstein05,Dolgopolov15,Kukushkin15,Shashkin02}. In this
case as well, we suggest that the Stoner approach is at least qualitatively 
relevant (see Appendix for details). While we do obtain a  ferromagnetic
instability at a density slightly above the critical value where the carrier 
effective mass diverges, a proper theoretical treatment, including both
short- and long-range interactions from the beginning, is still missing.

We are now in a position to summarise our results in more detail, beginning 
with the 
low-density regime above the MIT, which is characterised by strong long-range
correlations. In Sect. \ref{sec:dolgo}, these were taken into account 
phenomenologically via effective mass renormalisation, Eq. (\ref{eq:dolgo}), 
as observed
experimentally\cite{Dolgopolov15} and predicted 
theoretically\cite{Finkelstein05}. While this mass renormalisation alone would lead to an increased magnetic
susceptibility\cite{Dolgopolov15}, we find that including the effects of 
on-site repulsion enhances susceptibility further, leading to a second-order
ferromagnetic transition. The latter takes place at a density which is a 
few per cent above the value corresponding to the (asymptotic) divergence
of the effective mass. This difference is relevant in the context of the
disagreement between presently available experimental 
results\cite{Kravchenko06,Kravchenko06-2,Reznikov03}. 
Further experiments are needed in order
to shed light on this controversy, and also to clarify whether the MIT 
corresponds to the effective mass divergence or (as we speculated) to the 
magnetic transition.

The origins of such a strong effect of short-range interactions at low 
densities become clear as we note [see Eq. (\ref{eq:a2d})] that, for example,
a five-fold increase in the effective mass $m_\bot$ has the same effect on 
the value of $k_Fa_{2D}$ as does the five-fold increase of $U_{2D}$ (for 
example, at $n=8 \cdot 10^{10}$ cm$^{-2}$, the value of $g$ would increase 
to $0.14$; we verified that a self-consistent change of $U_{2D}$ due to 
the increase of $m_\bot$ is negligible, as expected). Specifically, the
system even at $n \sim n_c$ becomes strongly interacting also in terms of
short-range interaction. We also remark that a strong short range interaction
can lead to strong renormalisation of the Fermi liquid parameters (including
an additional renormalisation of the effective mass), which was not taken 
into account in our work or elsewhere.
This highlights the need for a microscopic theory which
would include {\it both} long- and short-range interactions on the same footing.

In the metallic regime at high densities, where the long-range correlation 
effects become unimportant, the value of $ka_{2D}$ increases due
to the increasing $U_{2D}$ (see above). On the other hand, the wave functions 
begin to change under
the effects of an applied field (see Fig. \ref{fig:scheme}), as the mean field
energy scale $nU_{2D}$ becomes sufficiently large to perturb the transverse 
carrier motion. These two effects lead to a strong increase in magnetic 
susceptibility $\chi$ with $n$, ultimately resulting in a ferromagnetic 
transition. For our parameter values, this takes place at 
$n_{FM}\approx 1.15 \cdot 10^{14}$ cm$^{-2}$, which is presently beyond the
experimental range for a Si-MOSFET. However, this value was 
obtained (in Sect. \ref{sec:metal}) without taking into account the Fermi liquid
renormalisations (such as effective mass enhancement, cf. Refs. 
\cite{Randeria,Randeria2,Bloom72}),
which again become important in this regime and may lower the 
value of critical density. Beyond mean-field description, fluctuation 
effects\cite{Conduit2013} may lead to a further decrease of this quantity.

As explained above, the wave functions change under the effect of an applied 
field leads to relaxing the Stoner criterion of ferromagnetism. In terms
of critical density, this means that the obtained value of $n_{FM}$ is lowered
in comparison to naive Stoner-based estimates (which are invalid in the case 
of geometrically restricted systems such as inversion layer). In addition, this
gives rise to a non-linear field dependence of magnetisation. The latter was
discussed previously for the case of quasi-2D systems with Coulomb 
interaction\cite{Tutuc03,DasSarma2006,Subasi2008}, albeit at smaller
densities, and our results outlined in Sect. \ref{sec:field} thus provide an 
additional mechanism for such non-linearity. 

Whether the actual high-density ferromagnetic transition is reachable or not,
the minimum and the subsequent increase of $\chi$ with density at 
$n \gtrsim 2 \cdot 10^{12}$ cm$^{-2}$ should be
observable. We also note that the threshold density 
$n^{(0,1)}\approx 3.3 \cdot 10^{12}$ cm$^{-2}$, beyond which the second transverse level is populated at $H=0$, is not far from the highest value used in 
the measurements to date   ($n =2.08 \cdot 10^{12}$ cm$^{-2}$, see 
Ref. \cite{epl12}), and should be attainable experimentally. 
In addition to new and potentially interesting transport phenomena arising at 
this point, one should be able
to measure the associated jump in the susceptibility $\chi$ 
(cf Fig. \ref{fig:chimulti}). 
With the help of Eq. (\ref{eq:2dstonermulti}), this can be used to calibrate 
$U_{2D}^{i,j}$, and ultimately $U$. Note that the value of $n^{(0,1)}$ can be 
further reduced
by decreasing the acceptor density $N_A$.

In order to keep our description simple, we omitted a number of effects which
are expected to be of quantitative importance only. These include a more
accurate formulation of the wave-function boundary conditions at $z=0$ 
(Ref. \cite{Stern72}), the image-charge potential\cite{Ando82}, etc.
Significantly, we also disregard the effects of the valley 
degree of freedom, where an accurate description would involve using the 
appropriate values (not yet available) for the strength of short-range 
interaction between the carriers belonging to  different valleys.  
Note that once such more accurate model is constructed, the important issue
of valley ``polarisation''\cite{Renard14} can be treated in the same way as 
that of spin
polarisation.
% using a similar approach.

In the present work, we specifically aimed at describing 
Si-(100) MOSFETs, however our results are expected to be qualitatively
relevant for other 2D electron systems of finite thickness. These general
conclusions are: (i) At higher densities, proper treatment requires taking 
into account the wave function change under the applied in-plane magnetic 
field\footnote{In principle, a similar wave function change should occur 
in various geometrically restricted systems, 
including quantum dots where it would lead to a magnetisation dependence of
electron interaction energies (including exchange). While this would be 
relevant for 
the studies of magnetic properties of quantum dots (cf. Ref. 
\cite{Burmistrov,Burmistrov2}),
the effect might prove negligible owing to 
the large quantisation energies.}    
(see Fig. \ref{fig:scheme}). This
effect leads to an increased susceptibility in the paramagnetic state and 
enhances
the tendency toward ferromagnetism. (ii) When the long-range correlations at 
low densities
lead to the effective mass enhancement (as in Si-MOSFET\cite{Dolgopolov15} 
or in 
GaAs quantum wells\cite{Kukushkin15}),
magnetic properties 
are significantly affected by the on-site carrier repulsion, which can lead
to a ferromagnetic instability.

\begin{acknowledgements}
The author takes pleasure in thanking R. Berkovits, P. Coleman, 
B. D. Laikhtman, S. V. Kravchenko, I. Shlimak,  L. D. Shvartsman, and R. Valenti 
for enlightening discussions. Discussions with the late K. A. Kikoin are 
gratefully acknowledged. This work was supported by the Israeli 
Absorption Ministry.
\end{acknowledgements}

\begin{appendices}

%\appendix{On the Applicability of Stoner-Type Mean Field Approach in Low-Density
%2D Systems }
\numberwithin{equation}{section}

\section{\hspace{-2.8mm} PPENDIX: On the applicability of Stoner-type mean field approach in low-density
2D systems }

In this work, we consider low-density (quasi-)2D electrons, and one might ask 
whether the short-range repulsion can affect the properties of the system
 in our range of values of 
parameters. If the answer were in the negative, this would have turned our 
mean-field treatment into an artifact of an inadequate approach. It is 
therefore important to consider this issue in more detail (in addition to
discussing the scattering length in Sect. \ref{sec:conclu}).

For simplicity, we consider a purely 2D system, 
\begin{equation}
{\cal H}= \sum_{i}\frac{\vec{p}^{\,2}_i}{2 m_\bot}+\frac{1}{2} \sum_{i\neq j} U_{2D} \delta(\vec{r}_i- \vec{r}_j)\,,
\end{equation}
where the summations are over the particle numbers. The effective 2D interaction
 $U_{2D}$ is in our case given by Eq. (\ref{eq:U2D});  calculations
 of Sect. \ref{sec:single} (cf. Fig. \ref{fig:U2D}) yield the value of $U_{2D}\approx $ 
1.2$\cdot$10$^{-27}$ erg$\cdot$cm$^{2}$  at
$n = 8 \cdot 10^{10}$ cm$^{-2}$ and  $U_{2D}\approx $5.3$\cdot$10$^{-27}$ erg$\cdot$cm$^{2}$  at $n=$8$\cdot$10$^{13}$ cm$^{-2}$. The level indices are 
suppressed as presently we are considering the single-level case.
Now let us consider interaction of a sole spin-down electron 
%(assumed to be at rest for simplicity) 
with the spin-up Fermi sea. The mean-field result for
the net interaction energy is of course $\delta E_{mf}=U_{2D} n_\uparrow$ (where 
at $M=0$, $n_\uparrow=n/2$), 
and our worry is that this expression may be a gross over-estimate. 
Indeed, with increasing $U_{2D}$ spin-up electrons will be avoiding the site occupied
by the spin-down electron, resulting in a smaller energy change which retains
a finite value $\delta E_{\infty}$  (of the order of the Fermi energy or less) 
even as $U_{2D}$ increases to infinity. The situation may arise where actually
\begin{equation}
\delta E_{\infty} < \delta E_{mf}=U_{2D} n/2 ,
\label{eq:invalid}
\end{equation}
in which case we suspect that the mean field estimates become irrelevant. Note
that in reality there is a finite concentration of spin-down particles and
the perturbations of spin-up Fermi sea by individual spin-down electrons are
not independent, so that $n_\downarrow \delta E_{\infty}$ {\it underestimates} 
the interaction energy at large $U_{2D}$.
   In order to estimate $\delta E_{\infty}$, we first evaluate the energy 
change $\Delta E$ of
a spinless two-valley ideal 2D Fermi gas (${\cal H}_0=\vec{p}^2/2m_\bot$) under 
the perturbing effect of a static impurity at origin 
[corresponding to potential energy ${\cal V}=V\delta(\vec{r})$]. 
Using the Lifshits--Krein trace formula\cite{LK}, this is
conveniently expressed as an integral from the bottom of the band to the 
Fermi energy,
\begin{equation}
\Delta E (V)= 2\int_0^{\epsilon_F} \xi(\epsilon) d \epsilon\,.
\label{eq:DE}
\end{equation}
Here, the prefactor corresponds to the two independent valleys, and the 
spectral shift function $\xi$ [with the property that  $-d \xi/d \epsilon$ 
equals $\delta \nu(\epsilon)$, an impurity-induced correction to the density of 
states $\nu(\epsilon)$] is given by\cite{LK,LK2,LK3}
\begin{eqnarray}
\xi(\epsilon)&=&-\frac{1}{\pi}{\rm Arg} {\rm Det} \left\{\hat{1} -\frac{1}
{\epsilon - {\rm i} 0- {\cal H}_0}{\cal V} \right\}= 
\nonumber \\ 
& =&-\frac{1}{\pi}{\rm Arg} \left\{{1} -V \int\frac{d^2k}{4 \pi^2} \frac{1}
{ \epsilon  - {\rm i} 0- (k^2/2 m_\bot)}\right\}= 
\nonumber \\ 
&=&-\frac{1}{\pi}{\rm Arg} \left\{{1} -V \dashint_0^W \frac{\nu(\epsilon') 
d \epsilon'}{\epsilon - \epsilon'} -\pi {\rm i} V \nu(\epsilon) \right\}\,.
\end{eqnarray}
Here, the momentum integral is over the Brilloin zone, whereas the energy 
integral in the last line is over the entire band, $0< \epsilon' < W$.

Since we will ultimately need to integrate $\xi$, the weak singularity at 
$\epsilon=0$ is unimportant. In the low-density case of $\epsilon_F \ll W$ 
we estimate
\[
\dashint_0^W \!\!\!\frac{\nu(\epsilon') 
d \epsilon'}{\epsilon - \epsilon'} \sim \nu_0 \log\!\! 
\left(\!\frac{\epsilon}{W} 
\right) \sim  \nu_0 \log\!\! \left(\!\frac{\epsilon_F}{W} \right) 
\sim \nu_0 \log
\!\!\left(\!\frac{n_\uparrow}{2N_0}\right)\,\!,
\]
with $\nu_0$ given by Eq. 
(\ref{eq:nu0}), and $N_0 \sim 1/a^2$ (where $a$ is the lattice period), 
the full capacity
of the 2D band for fixed spin and valley indices. Thus, we find
\begin{equation}
\xi(\epsilon) \approx \frac{1}{\pi} {\rm arc} \tan \frac{\pi V \nu_0}{1-V \nu_0 \log(n_\uparrow a^2/2)}\,.
\label{eq:xi}
\end{equation}
Spectral shift function is related\footnote{For a recent mathematical 
discussion, see Ref. \cite{Kohmoto13}.}  by the Friedel sum rule 
to the scatterring phase shift\cite{iml48}, with the Born 
approximation corresponding to omitting the logarithmic term in Eq. 
(\ref{eq:xi}).
At small $V$, Eq. (\ref{eq:DE}) then yields the expected perturbative result
$\Delta E = V n_\uparrow$, whereas for large $V \stackrel{>}{\sim} 1/|\nu_0 
\log(n_\uparrow a^2/2)|$ we find
\begin{equation}
\Delta E(\infty) \approx 2 \epsilon_F \left|\log(n_\uparrow a^2/2)\right|^{-1}
\,.
\end{equation}
The latter is the energy change of spin-up Fermi sea when a node at 
$\vec{r}=0$ is created in all the electron wave functions. It is seen that 
indeed at very low densities $\Delta E(\infty)/\epsilon_F$ vanishes 
logarithmically,
which is the physical reason why the short-range interaction becomes 
irrelevant at sufficiently low densities. In our case, however, the 
absolute value of the
log does not exceed 10.
 
  In addition, note that the quantity $\delta E_\infty$  involves interaction 
with a spin-down electron which is not localised at origin but is moving with a 
velocity of order of $v_F$. The wave-functions node is presumably a heavy 
object, and moving it along would result in a large addition to 
$\Delta E(\infty)$.
It is thus more economical to have the spin-down electron localised in an 
area of size $R \sim \hbar/p_F$ (which can be done without appreciably changing 
its energy) while requiring that the wavefunctions of the spin-up electrons 
vanish throughout this area. The corresponding energy change of the spin-up 
Fermi sea is a sum of $\Delta E(\infty)$ and an area term, needed to 
``inflate'' the node to the required finite area: 
\begin{eqnarray}
\delta E_\infty \sim \Delta E(\infty)+2 \nu_0 \int_0^{\epsilon_F} \epsilon 
d\epsilon R^2 = \nonumber \\
= 2 \epsilon_F \left|\log(n_\uparrow a^2/2)\right|^{-1} + \frac{1}{4\pi} 
\epsilon_F\,.
\label{eq:DEinf}  
\end{eqnarray}
Throughout our range of parameter values, the second term is at least several 
times
smaller than the first one, hence we do not need a more elaborate estimate
of the energy of correlated motion of spin-down electron. We are now in a 
position to quantitatively verify that we never approach the ``dangerous'' 
regime
specified by the inequality (\ref{eq:invalid}). Since presently we did not 
take into account the possibility of multiple occupied subbands (which is 
not expected to qualitatively affect the
results), this must be done with the help of the numerical results obtained 
for the single-level case, Sect. \ref{sec:single}. 

Using the values of $U_{2D}$ quoted above, we find that at 
$n =8\cdot 10^{10}$ cm$^{-2}$ (where the Fermi energy as measured form the bottom
of the band is $\epsilon_F \equiv E_v-E_0\approx 0.51$ meV),  the value of 
$\delta E_{mf} \approx 0.029$ meV is about 5 times smaller than $\delta E_\infty \approx 0.15$ meV. 
Likewise, at  
$n=8\cdot 10^{13} $cm$^{-2}$ (where $\epsilon_F \approx$ 0.51 eV), the value 
of  $\delta E_{mf} \approx 132$ meV is smaller than  
$\delta E_\infty \approx 400$ meV.

We thus conclude that the mean-field estimate of the interaction energy, and by
extension the Stoner approach, should be at least qualitatively applicable 
throughout the entire range of densities considered herein. Since a 
Stoner-type treatment is anyhow not expected to be quantitatively accurate, 
this is a satisfactory outcome.

One further note should be made concerning the situation at very low densities 
near MIT (Sect. \ref{sec:dolgo}). In this case, the long-range forces lead to 
a significant reduction of effective band width (and hence of the effective
Fermi energy), to the extent that if those renormalised quantities are 
substituted when calculating $\delta E_\infty$, one might find that the 
inequality (\ref{eq:invalid}) is actually satisfied. We wish to argue that 
such a substitution would be hard to justify, quoting the following reasons:

\noindent (i) the renormalised quantities refer not to the electrons, but 
to the 
resultant quasiparticles. These are extended objects, which presumably should 
be viewed as residing  on an effective lattice with proportionally increased 
lattice period, which should thus be used in place of $a$ in Eq. 
(\ref{eq:DEinf}).

\noindent (ii) More importantly, these quasiparticles characterise low-energy, 
long-wavelength properties of the system, whereas contact interaction with 
point defects involves a significant short-wavelength component. The 
short-wavelength
contribution to Eq. (\ref{eq:DE}) originates from the logarithmic term
in Eq. (\ref{eq:xi}). Therefore, it is more appropriate to use 
{\it unrenormalised} spectral parameters when estimating this term {\it only}, 
including the coefficient before the logarithm. Elsewhere in Eqs. (\ref{eq:DE})
and (\ref{eq:xi}), one should be using the renormalised spectrum characterised
by a larger mass, yet it is easy to see that within this order-of-magnitude
estimate the
renormalisation coefficient cancels out for large $V$.
Hence $\delta E(\infty)$ retains (roughly) its unrenormalised value and we 
arrive at a 
conclusion that the mean-field approach is
still qualitatively applicable.      
\end{appendices}

\end{document}